\documentclass[american,aps,pra,reprint,superscriptaddress, longbibliography,floatfix]{revtex4-1}
\usepackage[T1]{fontenc}
\usepackage{inputenc}
\usepackage{color}
\usepackage{babel}
\usepackage{physics}
\usepackage{mathtools}
\usepackage{bbding}
\usepackage{pifont}
\usepackage{textcomp}
\usepackage{wasysym}
\usepackage{amsthm}
\usepackage[normalem]{ulem}
\usepackage{nicefrac}
\usepackage{wasysym}
\usepackage{dsfont}
\usepackage{amsmath}
\usepackage{amssymb}
\usepackage{graphicx}
\usepackage {hyperref}

\hypersetup{
     colorlinks   = true,
     linkcolor    = blue,
     citecolor    = blue,
     urlcolor     = blue,
     }

\usepackage{appendix}
\renewcommand{\selectlanguage}[1]{}
\makeatletter
\@ifundefined{textcolor}{}
{%
	\definecolor{BLACK}{gray}{0}
	\definecolor{WHITE}{gray}{1}
	\definecolor{RED}{rgb}{1,0,0}
	\definecolor{GREEN}{rgb}{0,1,0}
	\definecolor{BLUE}{rgb}{0,0,1}
	\definecolor{CYAN}{cmyk}{1,0,0,0}
	\definecolor{MAGENTA}{cmyk}{0,1,0,0}
	\definecolor{YELLOW}{cmyk}{0,0,1,0}
}
\theoremstyle{plain}

\theoremstyle{plain}

\ifx\proof\undefined
\newenvironment{proof}[1][\protect\proofname]{\par
	\normalfont\topsep6\p@\@plus6\p@\relax
	\trivlist
	\itemindent\parindent
	\item[\hskip\labelsep
	\scshape
	#1]\ignorespaces
}{%
	\endtrivlist\@endpefalse
}
\providecommand{\proofname}{Proof}
\fi
\theoremstyle{plain}

\providecommand{\lemmaname}{Lemma}
\providecommand{\definitionname}{Definition}
\providecommand{\propositionname}{Proposition}

\usepackage{babel}
\usepackage{txfonts}%\usepackage{braket}
\usepackage{colortbl}\definecolor{myurlcolor}{rgb}{0,0,0.7}

\def\ket#1{| #1 \rangle}
\def\bra#1{\langle  #1 |}

\newcommand{\haH}

\definecolor{orange}{RGB}{255,127,0}

\begin{document}

\title{Coherent heat transfer leads to genuine quantum enhancement in the
performances of continuous engines}

\author{Brij Mohan}
\email{brijhcu@gmail.com}
\affiliation{Department of Physical Sciences, Indian Institute of Science Education and Research (IISER), Mohali, Punjab 140306, India}

\author{Rajeev Gangwar}
\affiliation{Department of Physical Sciences, Indian Institute of Science Education and Research (IISER), Mohali, Punjab 140306, India}

\author{Tanmoy Pandit}
\affiliation{Fritz Haber Research Center for Molecular Dynamics, Hebrew University of Jerusalem, Jerusalem 9190401, Israel}

\author{Mohit Lal Bera}
\affiliation{Departamento de F\'{i}sica Te\'{o}rica and IFIC,  Universidad de Valencia-CSIC, 46100 Burjassot (Valencia), Spain}
\affiliation{ICFO - Institut de Ci\`encies Fot\`oniques, The Barcelona Institute of Science and Technology, 08860 Castelldefels (Barcelona), Spain}

\author{Maciej Lewenstein}
\affiliation{ICFO - Institut de Ci\`encies Fot\`oniques, The Barcelona Institute of Science and Technology, 08860 Castelldefels (Barcelona), Spain}
\affiliation{ICREA, Pg. Lluis Companys 23, ES-08010 Barcelona, Spain}

\author{Manabendra Nath Bera}
\email{mnbera@gmail.com}
\affiliation{Department of Physical Sciences, Indian Institute of Science Education and Research (IISER), Mohali, Punjab 140306, India}

\begin{abstract}
Conventional continuous quantum heat engines with incoherent heat transfer perform poorly as they exploit two-body interactions between the system and hot or cold baths, thus having limited capability to outperform their classical counterparts. We introduce distinct continuous quantum heat engines that utilize coherent heat transfer with baths, yielding genuine quantum enhancement in performance. These coherent engines consist of one qutrit system and two photonic baths and enable coherent heat transfer via two-photon transitions involving three-body interactions between the system and hot and cold baths. We demonstrate that coherent engines deliver significantly higher power output with much greater reliability, i.e., lower signal-to-noise ratio of the power, by hundreds of folds over their incoherent counterparts. Importantly, coherent engines can operate close to or at the maximal achievable reliability allowed by the quantum thermodynamic uncertainty relation. Moreover, coherent engines manifest more nonclassical features than incoherent engines because they violate the classical thermodynamic uncertainty relation by a greater amount and for a wider range of parameters. These genuine enhancements in the performance of coherent engines are directly attributed to their capacity to harness higher energetic coherence for the resonant driving case. The experimental feasibility of coherent engines and the improved understanding of how quantum properties can enhance performance may find applications in quantum-enabled technologies.

\end{abstract}

\maketitle
\section{introduction}
Quantum heat engines -- microscopic thermal devices designed to convert heat into quantum mechanical work -- have become one of the focal points of research considering the current quantum industrial revolution \cite{Binder2018, Alexia2022, Mukherjee2024}. This leads to studying thermodynamics in the microscopic and quantum regime, both from foundational and applied aspects \cite{Jarzynski1997, Crooks1999, Campisi2011, Brando2013, Horodecki2013, Skrzypczyk2014, Brando2015, Lostaglio2015, Alhambra2016, Bera2017, Sparaciari2017, Binder2018, Aberg2018, Gour2018, Muller2018, Uzdin2018, Bera2019, Bera2021,  Bera2022, Khanian2023, Bera2024}. The earliest model of a quantum heat engine was proposed by Scovil and Schulz-DuBois (SSD), which is composed of a qutrit interacting with two thermal baths \cite{Scovil1959}. Later, it was re-investigated in a full quantum setting using open quantum system dynamics~\cite{ Boukobza2006b, Boukobza2006a, Boukobza2007, Kosloff2014}. In the last decades, many other models of quantum heat engines have been proposed; see Refs.~\cite{Kosloff2014, Binder2018, Myers2022, Cangemi2023} for a comprehensive overview of historical and recent advancements. Optomechanical systems~\cite{Sheng2021}, nitrogen-vacancy centers in diamond~\cite{Klatzow2019}, trapped ions~\cite{Robnagel2016, Bouton2021}, nuclear magnetic resonance (NMR)~\cite{Peterson2019}, and superconducting circuits~\cite{Clodoaldo2023} have emerged as versatile experimental platforms to realize quantum heat engines, bringing these theoretical concepts into practical realizations.

The conventional continuous quantum heat engines operate in a steady-state regime, by interacting continuously with hot and cold baths~\cite{Kosloff2014, Binder2018, Myers2022, Cangemi2023}. These engines, in general, deliver low power with high fluctuation~\cite{Kosloff2014, Rahav2012, Kalaee2021, Pena2021, Singh2023}. As a result, the reliability, i.e., the ratio between the variance and average of power (or relative fluctuation in power), of these engines is considerably compromised. 
It has been observed that continuous quantum thermal devices, when energetic coherence is present, may enhance power~\cite{Scully2011, Um2022} and efficiency~\cite{Scully2010, Dorfman2018}, improves reliability (less relative fluctuation in power)~\cite{Kalaee2021, Singh2023}, and may lead to violation of classical thermodynamic trade-off relations (classical thermodynamic uncertainty relation (cTUR)~\cite{Barato2015} and power-efficiency-constancy trade-off relation~\cite{Pietzonka2018})~\cite{Ptaszy2018, Liu2019, Pal2020, Rignon2021, Pena2021, Kalaee2021, Van2022, Souza2022, Singh2023, Prech2023, Manzano2023, Jaseem2023}. These violations indicate that these engines can operate in the quantum regime. However, it does not necessarily imply that quantum engines are operating close to their optimal capacity in terms of reliability. Ideally, one would expect negligible relative fluctuation in power from an ideal continuous engine. However, relative fluctuation cannot be suppressed to zero due to the existence of a finite lower bound on it determined by the quantum thermodynamic uncertainty relation (qTUR)~\cite{Hasegawa2020}. This lower bound represents a fundamental quantum limit, which is derived from the celebrated quantum Cramér-Rao bound~\cite{Braunstein1994}, and is closely related to the so-called quantum speed limits~\cite{Hasegawa2020, Hasegawa2023}.

The characteristic feature of traditional continuous quantum heat engines is that they utilize incoherent heat transfers between the working system and the baths. It implies that the transitions induced in the working system by the hot and cold baths are independent (or uncorrelated), rendering them highly stochastic in nature. This feature constitutes one of the reasons for these engines to have limited ability to outperform their classical counterparts. For the in-depth comparison of continuous thermal machines and their classical counterparts, please see Refs.~\cite{Kalaee2021, Almanza-Marrero2024}. Therefore, we are required to reduce the stochastic nature of the transitions in the working system induced by the baths to overcome these limitations. The natural question is, thus, how to employ an operationally distinct heat transfer mechanism, rather than the incoherent one, in continuous heat engines that inherently involve less stochastic transitions and lead to significant enhancement in performance.

In this article, we affirmatively address the above question by introducing the concept of a coherent heat transfer mechanism in continuous heat engines in which the baths induce correlated (or mutually dependent) transitions in the working system, and, as a result, the stochastic nature of transition decreases. The continuous engines operating with this mechanism are termed coherent quantum heat engines (CQHEs). These engines can be physically realized by considering a qutrit coherently interacting with hot and cold baths through two-photon transitions (Raman interaction, i.e., three-body interactions between system and baths) in the presence of periodic driving by an external field. The analogous incoherent quantum heat engines (IQHEs) are the standard SSD engines~\cite{ Boukobza2006b, Boukobza2006a, Boukobza2007, Kosloff2014}, where a qutrit interacts incoherently (independently, through one-photon transitions) with the hot and cold baths. For the same set of qutrit and bath parameters, the CQHEs deliver much higher power and much lower relative fluctuation in power compared to IQHEs  with both resonant and non-resonant driving. In fact, the performance of CQHEs can be enhanced by hundreds of folds of that of IQHEs. This enhancement is directly attributed to the presence of a much higher amount of energetic coherence in CQHEs with resonant driving, which is a consequence of coherent heat transfer. Moreover, for the same reason, the CQHEs not only exhibit a more profound violation of cTUR and power-efficiency-constancy trade-off relations compared to IQHEs but also can suppress relative fluctuation in power to the quantum limit imposed by qTUR. Hence, CQHEs manifest genuine quantum enhancement over IQHEs and classical engines.

The rest of the article is organized as follows. In section~\ref{sec:CQHE}, we introduce the generic models of continuous quantum coherent and incoherent engines involving coherent and incoherent heat transfers, respectively. We demonstrate the genuine quantum enhancements in performances by coherent engines over incoherent engines in section~\ref{sec:QuantAdvs}. Finally, our results are summarized in section~\ref{sec:Summary}.

\section{Continuous coherent quantum heat engines \label{sec:CQHE}}
A continuous heat engine consists of a working system that weakly interacts with two heat baths at different temperatures while, at the same time, being periodically driven by an external field. The simplest model for such an engine utilizes a qutrit system interacting with two baths, widely studied in literature \cite{Scovil1959, Boukobza2006b, Boukobza2006a, Boukobza2007, Kosloff2014, Myers2022, Cangemi2023}. Explicitly, a qutrit with Hamiltonian $H_{S}=\omega_{h} \ketbra{2} + (\omega_{h}-\omega_{c})\ketbra{1}$ is coupled to two thermal (photon) baths with respective inverse temperatures $\beta_c$ and $\beta_h$, where $\omega_h > \omega_c$ and $\beta_c > \beta_h$. In addition, the qutrit is driven by an external field following the Hamiltonian $H_d(t)= \alpha(e^{-i\omega_d t}\ketbra{1}{0}+e^{i\omega_d t}\ketbra{0}{1})$. The condition $\beta_h \omega_h < \beta_c \omega_c$ needs to be ensured for this device to operate as a heat engine {{(see Ref.~\cite{Kosloff2014} and Appendix~\ref{appsec:CQHEsteaystate})}}. We assume $\hbar = k_B = 1$ throughout this work. The total Hamiltonian of the qutrit-baths composite is
\begin{align*}
H=H_{S}(t) + H_{B_h} + H_{B_c} + H_{SB_hB_c}^X,    
\end{align*}
where $H_{S}(t) = H_{S} + H_d(t)$ is the total Hamiltonian of the qutrit, $H_{B_h}=\sum_k \Omega_{k,h} \ a^\dag_{k,h} a_{k,h}$ and $H_{B_c}= \sum_{k'} \Omega_{k',c} a^\dag_{k',c} a_{k',c}$ are the Hamiltonians of the hot and cold (photon) baths with mode frequencies $\Omega_{k,h}$ and $\Omega_{k',c}$ respectively, and $H_{SB_hB_c}^X$ represents the interaction between the qutrit and the baths. 

\begin{figure}
\centering    
\includegraphics[width=7.5cm,height=4.5cm]{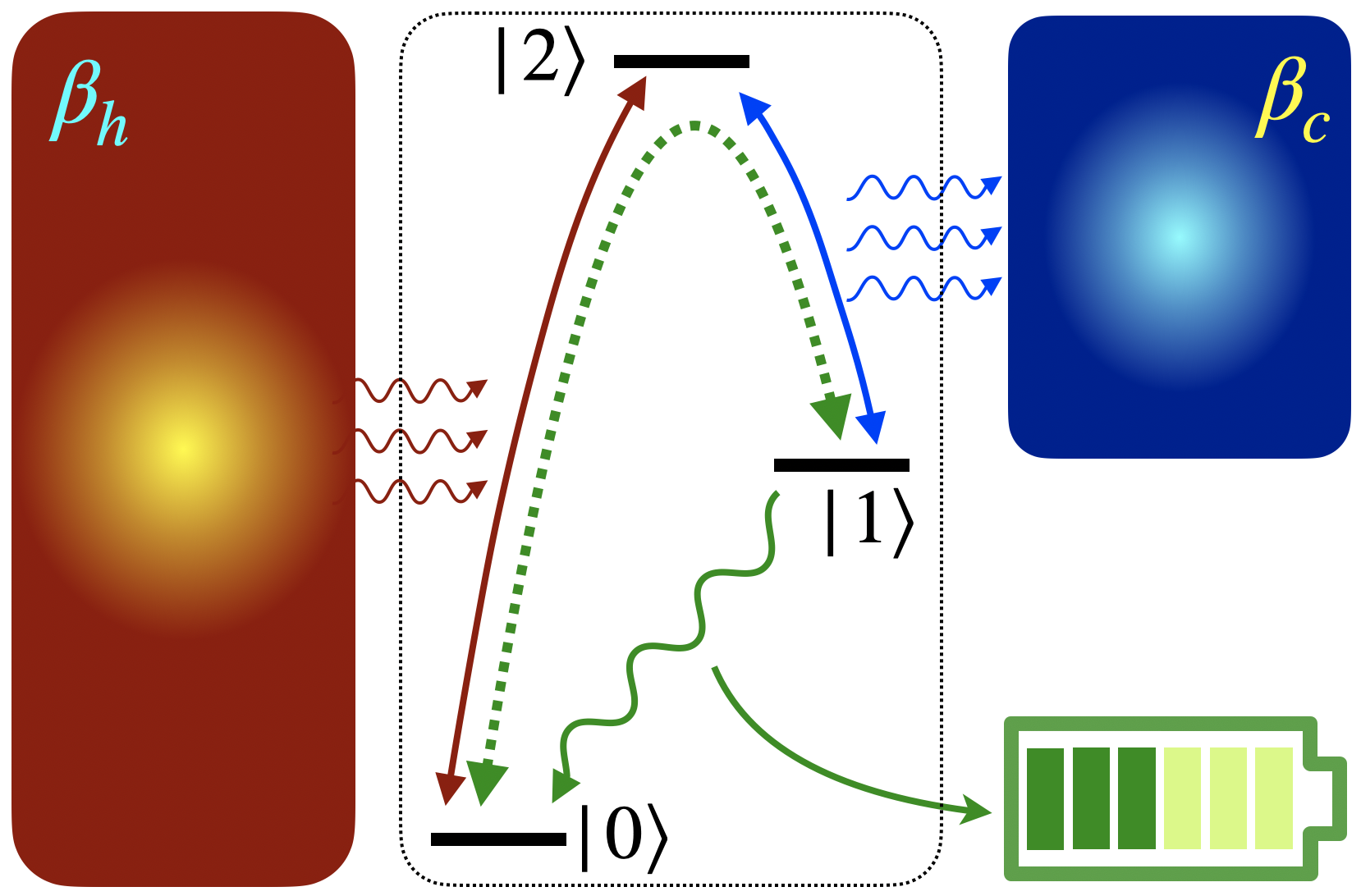}
\caption{ {\bf Schematics of incoherent and coherent heat engines.} The engine is constituted by a three-level quantum system (qutrit), which weakly interacts with hot and cold baths with the inverse temperatures $\beta_h$ and $\beta_c$. In incoherent heat engine, the energy (heat) transfer takes place via (independent) single photon transitions, i.e., energy levels $\ket{0}$ and $\ket{2}$ interact with the hot bath and levels $\ket{1}$ and $\ket{2}$ interact with the cold bath, governed by the interaction Hamiltonian~\eqref{eq:IntHamInCoh}. Solid (red and blue) arrows indicate these independent or incoherent energy transfers. In coherent heat engines, the energy transfer takes place via two-photon transitions, where effectively energy levels $\ket{0}$ and $\ket{1}$ participate in the process, and absorption of a photon from the hot bath is associated with the release of a photon top the cold bath and vice versa. This coherent heat transfer is governed by the interaction Hamiltonian~\eqref{eq:IntHamCoh} and indicated here by the dotted (green) arrow. The wavy arrow (solid-green) between $\ket{0}$ and $\ket{1}$ indicates the external driving. See text for more details. }
\label{fig:CAIQHE}
\end{figure}

Below, we consider two qualitatively different models of continuous heat engines that differ in the interaction between the qutrit and the baths, i.e., $H_{SB_hB_c}^X$. In particular, our goal is to compare the performances of engines with an interaction Hamiltonian ($H_{SB_hB_c}^C$) that only allows `coherent' energy transfer with the performances of engines with an interaction Hamiltonian ($H_{SB_hB_c}^I$) that enables `incoherent' energy transfer between the baths and the qutrit. See the schematics of coherent and incoherent engine interactions given in Fig.~\ref{fig:CAIQHE}. \\

{\it Coherent Quantum Heat Engines (CQHEs)} -- We introduce an engine that involves energy transfer between the baths and the qutrit via a two-photon process, driven by an interaction Hamiltonian~\cite{Gerry1990,Gerry1992,Wu1996}
\begin{align}
H_{SB_hB_c}^C= g_0 \sum_{k,k'} (a_{k,h} a_{k',c}^\dag b_{hc}^\dag + a_{k,h}^\dag a_{k',c} b_{hc}), \label{eq:IntHamCoh}  
\end{align}
where $b_{hc}=\ketbra{0}{1}$ and $g_0$ is the coupling strength. Here, the energy transfer between the baths and the system is {\it coherent} in the sense that any photon absorbed from the hot bath is associated with a release of a photon to the cold bath and the excitation $\ket{0} \to \ket{1}$, and vice versa. For $|g_0| \ll 1$, the local dynamics of the qutrit reduces to
\begin{align}
    \dot{\rho} = i \ [\rho, \ H_{S}(t)] + \mathcal{D}_{hc} (\rho)
\end{align}
for a qutrit state $\rho$, where the only dissipator in the Lindblad master equation is given by,
\begin{align*}
    \mathcal{D}_{hc}(\rho) &=  \gamma_{1} (b_{hc} \rho b_{hc}^{\dag}  -\{b_{hc}^{\dag}b_{hc},\rho\}/2)  \nonumber \\
    &+ \gamma_{2} (b_{hc}^{\dag} \rho b_{hc} - \{b_{hc} b_{hc}^{\dag},\rho\}/2),
\end{align*}
with $\gamma_1=\gamma_0 n_c(n_h+1)$, $\gamma_2=\gamma_0n_h(n_c+1)$, and $\gamma_0$ is Weiskopf-Wigner decay constant. The derivation of the above Lindblad master equation is outlined in (see Appendix~\ref{appsec:CQHEsteaystate}). The dissipator $\mathcal{D}_{hc}$ involves the parameters of both hot and cold baths and induces dissipation utilizing the levels $\ket{0}$ and $\ket{1}$. The level $\ket{2}$ is never ``engaged'' in the process. Due to the coherent nature of the interaction, the energy (heat) transfer between the baths and the qutrit is less random (i.e., involves less stochastic transitions) due to correlated heat transfer than that of the engines with incoherent heat transfer considered earlier.

To calculate the power, heat currents, and other relevant quantities, we move to a rotating frame employing the transformation $B_{R}=e^{iH_Rt}Be^{-iH_Rt}$, where $B$ is an operator and $[H_{S}, H_{R}]=0$. With the resultant time-independent qutrit Hamiltonian $H_{rot}=H_S - H_R+H_{dR}$, where $H_{dR}=\alpha(\ketbra{1}{0}+\ketbra{0}{1})$, the dynamics attains a steady state in the rotating frame. For the steady state $\sigma_C$, with $\dot{\sigma}_C=0$, the average power $ \langle P_C \rangle $ is given by
\begin{align}
\langle P_C \rangle=-i \  \Tr([H_S, H_{dR}] \sigma_C) \leq 0.    
\end{align}
The average heat currents $\langle \dot{J}^x_C \rangle$ cannot be quantified directly because there are no independent dissipators corresponding to hot and cold baths. For that, we employ full counting statistics of the steady state dynamics (see Appendix~\ref{appsec:FCS}). This enables us to calculate the heat currents, the fluctuations in power ($\Delta P_C$), and the fluctuation in heat currents ($\Delta J^x_C$). With heat current from the hot bath $\langle \dot{J}^h_C \rangle$, we may compute the heat-to-work conversion efficiency $\eta_C=-\langle P_C \rangle /\langle \dot{J}^h_C \rangle $ of CQHEs.

\begin{figure*}
    \centering
    \includegraphics[width=7.8cm]{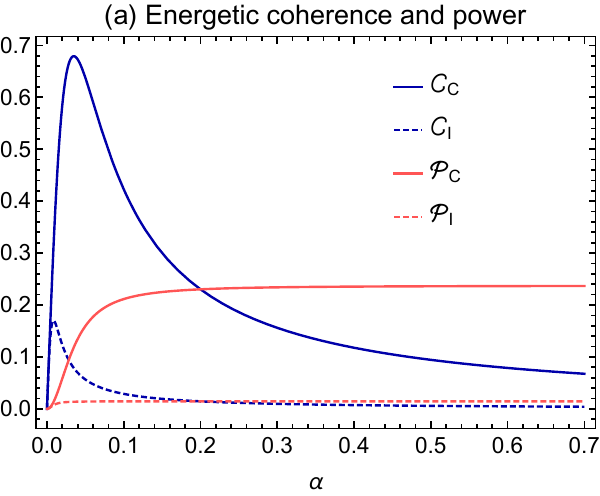}
    \space
\space
\space
     \includegraphics[width=9.5cm]{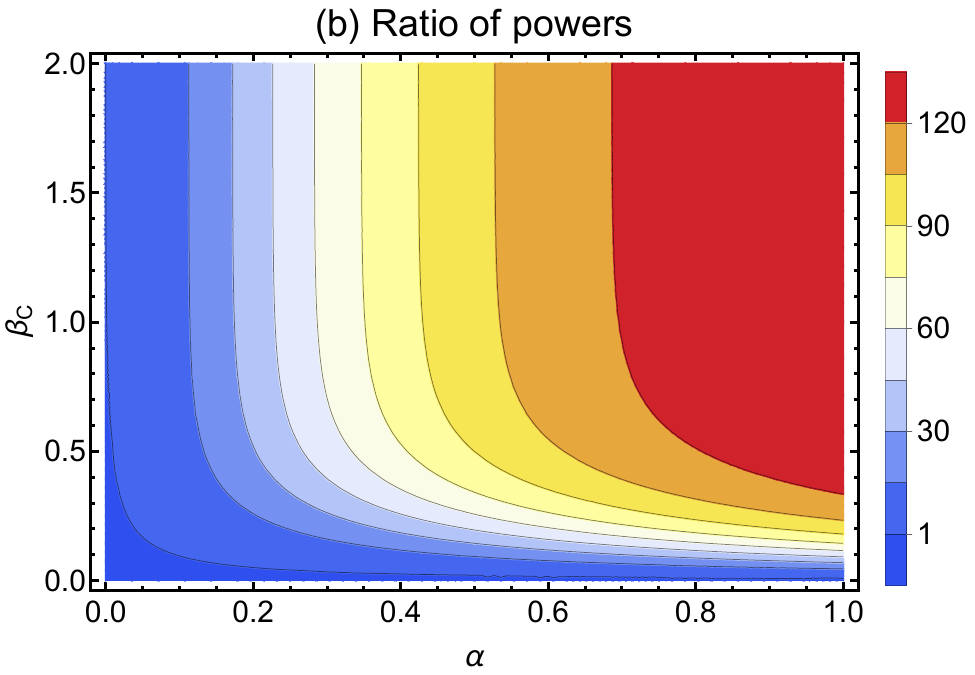}
\caption{{\bf Comparisons of energetic coherence and power outputs in coherent and incoherent engines.} The computations are carried out with the parameters $\gamma_{0} = 0.01$, $\omega_{h} = 10$, $\omega_{c} = 5$. (a) The figure on the left illustrates the variation in energetic coherence $\mathcal{C}_C=\mathcal{C}(\sigma_C)$ and $\mathcal{C}_I=\mathcal{C}(\sigma_I)$ for both coherent and incoherent heat engines, respectively with respect to the driving field strength $\alpha$, for $\beta_h=0.01$ and $\beta_c=0.8$. The expressions of energetic coherence are given Eqs.~\eqref{eq:CohCQHE} and~\eqref{eq:CohIQHE}. The traces in solid-blue and dashed-blue represent $\mathcal{C}_C$ and $\mathcal{C}_I$, respectively. The corresponding power outputs $\mathcal{P}_C$ and $\mathcal{P}_I$, given in Eq.~\eqref{eq:PtoC}, by coherent and incoherent engines, are presented with the solid-red and dashed-red traces, respectively. (b) The figure of the right displays the ratio of powers $\mathcal{P}_{C}/\mathcal{P}_{I} = \mathcal{C}_{C} / \mathcal{C}_{I}$ of the coherent and incoherent heat engine, with $\beta_h=0.001$, against $\alpha$ and $\beta_{c}$. In fact, for these parameters, the ratio can be $\mathcal{P}_{C} /\mathcal{P}_{I} \geq 135$. See text for more details.}
    \label{fig:CoherenceandPower}
\end{figure*}

{\it Incoherent Quantum Heat Engines (IQHEs)}---The conventional (continuous) quantum heat engines can be regarded as the incoherent analogs of CQHEs because they utilize incoherent energy transfers between working system and baths \cite{Boukobza2006b, Boukobza2006a, Boukobza2007} with the interaction Hamiltonian
\begin{align}
H_{SB_hB_c}^I= g_h \sum_k (a_{k,h} b^\dag_h + a_{k,h}^\dag b_{h}) + g_c \sum_{k'} (a_{k',c} b_{c}^\dag + a_{k',c}^\dag b_{c}), \label{eq:IntHamInCoh} 
\end{align}
where $b_{h} = \ketbra{0}{2}$ and $b_{c} =\ketbra{1}{2}$ are the ladder operator acting on the qutrit space. The coefficients $g_h$ and $g_c$ are the interaction strength with the hot and cold baths, respectively. The interaction drives incoherent energy (heat) transfer in the sense that the energy exchange between the states $\ket{0}$ and $\ket{2}$ with the hot bath is independent of the energy exchange between the states $\ket{1}$ and $\ket{2}$ with the cold bath. For $|g_h|, |g_c| \ll 1$, the local dynamics of the qutrit is expressed by the Lindblad master equation \cite{Breuer2007, Boukobza2006b, Boukobza2006a, Boukobza2007}, which contains two independent dissipators corresponding to hot and cold baths. The appearance of two dissipators in the Lindblad master equation reflects that the heat exchange with the hot bath is independent (or uncorrelated) of the heat exchange with the cold baths. Thus, the heat exchanges between the baths are incoherent. These incoherent heat engines are widely studied in literature and the remaining details are similar to CQHEs, for shake of completeness, see Appendix~\ref{InCEng}.

 The resultant time-independent total Hamiltonian of qutrit system for both coherent and incoherent engines in the rotating frame $H_{rot}= -\delta \op{1} +\alpha(\ketbra{1}{0}+\ketbra{0}{1})$, with detuning parameter $\delta=\omega_d-(\omega_h-\omega_c)$, where we have assumed $H_{R}=\omega_h\op{2}+ (\omega_d + (\omega_h-\omega_c))\op{1}$ without loss of generality (see Appendix~\ref{ROT}). In the next section, we analyze the performances of coherent and incoherent heat engines with resonant driving  ($\delta = 0$). However, we also discuss the non-resonant driving cases ($\delta \neq 0$) in Appendix~\ref{NonReso} for the shake of completeness.

\section{Quantum enhancements in coherent engines \label{sec:QuantAdvs}}
An evaluation of the performance of a continuous quantum heat engine requires a comprehensive analysis of three metrics: (i) efficiency, which signifies how efficiently heat is being converted into work; (ii) power, which is the rate of work output; and (iii) noise-to-signal ratio (NSR) in power, which signifies the relative fluctuation or inverse of precision in the power output. Here, we compare these metrics for coherent and incoherent heat engines with resonant driving and demonstrate that the former have substantial quantum enhancements in performance over the latter.

Our analysis reveals that the engine performance is related to the energetic coherence present in the steady state $\sigma_X$ (for $X=I, C$) in the rotating frame. Henceforth, a steady state refers to the steady state in the rotating frame. The quantum enhancements in the performance of CQHEs over the IQHEs are the direct consequence of the fact that the energetic coherence in $\sigma_C$ is higher than that of $\sigma_I$, in general. Note that the energetic coherence in the steady state results from a balance between two opposing processes - the (periodic) driving that creates coherence and the dissipation(s) that destroys coherence in the qutrit. Due to coherent heat transfer, the dissipative `tendency' in CQHEs is weaker compared to the dissipative `tendency' in IQHEs. As a result, we observe more energetic coherence in the former for the resonant driving case.

We start our analysis by studying the coherence in the steady states. In what follows, we set $\gamma_h=\gamma_c=\gamma_0$ and equal driving strength $\alpha$ for fair comparisons. The energetic coherence is measured using the $l$-1 norm of coherence ~\cite{Baumgratz2014}, given by $\mathcal{C}(\sigma_X)=\sum_{i\neq j}|\sigma_X^{(ij)}|$, where $\sigma_X^{(ij)}=\bra{i}\sigma_X \ket{j}$. For CQHEs and IQHEs with resonant driving, $\sigma_X^{(ij)} = \sigma_X^{(ji)*}$, and the corresponding amount of energetic coherence in the steady states are given by
\begin{align}
\mathcal{C}(\sigma_C)&=\frac{4 \alpha \   \gamma_{0} (n_{h}-n_{c})}{8 \alpha^2+\gamma_{0} ^2 (n_{hc}+2n_{c}n_{h})^2}, \label{eq:CohCQHE} \\ 
\mathcal{C}(\sigma_I)&=\frac{4 \alpha \  \gamma_{0}  (n_{h}-n_{c})}{4 \alpha^2 (3 n_{hc}+4)+ \gamma_{0} ^2 n_{hc} (n_{hc}+3 n_{h} n_{c})}, \label{eq:CohIQHE}
\end{align}
where $n_{hc}=n_h + n_c$. We refer to Appendices~\ref{appsec:IQHEsteaystate} and~\ref{appsec:CQHEsteaystate} for detailed derivation. For fixed $\gamma_0$, $n_h$, and $n_c$, the energetic coherence is a function of the driving strength $\alpha$. As shown in Fig.~\ref{fig:CoherenceandPower}(a), the energetic coherence $\mathcal{C}(\sigma_C)$ for CQHEs are higher than the energetic coherence $\mathcal{C}(\sigma_I)$ of IQHEs in general. Even for some reasonable values of system and bath parameters, the $\mathcal{C}(\sigma_C)$ becomes more than 135 times of $\mathcal{C}(\sigma_I)$, i.e., $\mathcal{C}(\sigma_C) \geq 135 \ \mathcal{C}(\sigma_I)$. We also note that, for fixed $n_h$, $n_c$, and $\gamma_0$, there is a threshold value of the driving strength $\alpha_{0}$ for which $\mathcal{C}(\sigma_C) = \mathcal{C}(\sigma_I)$. We calculate the threshold value (see Appendix~\ref{appsec:AlphaCrit}) and observe that $\mathcal{C}(\sigma_C) \leq \mathcal{C}(\sigma_I)$ for $\alpha \leq \alpha_{0}$. However, the $\alpha_{0}$ is generally very small, representing extremely weak periodic driving, except for the case of the baths with very high temperatures, i.e., $n_h \approx n_c \gg 1$. In all reasonable physical situations, the engines operate with $\alpha > \alpha_{0}$, which we consider for evaluating engine performances below. \\ 

\begin{figure*}
\centering
\includegraphics[width=6.2cm]{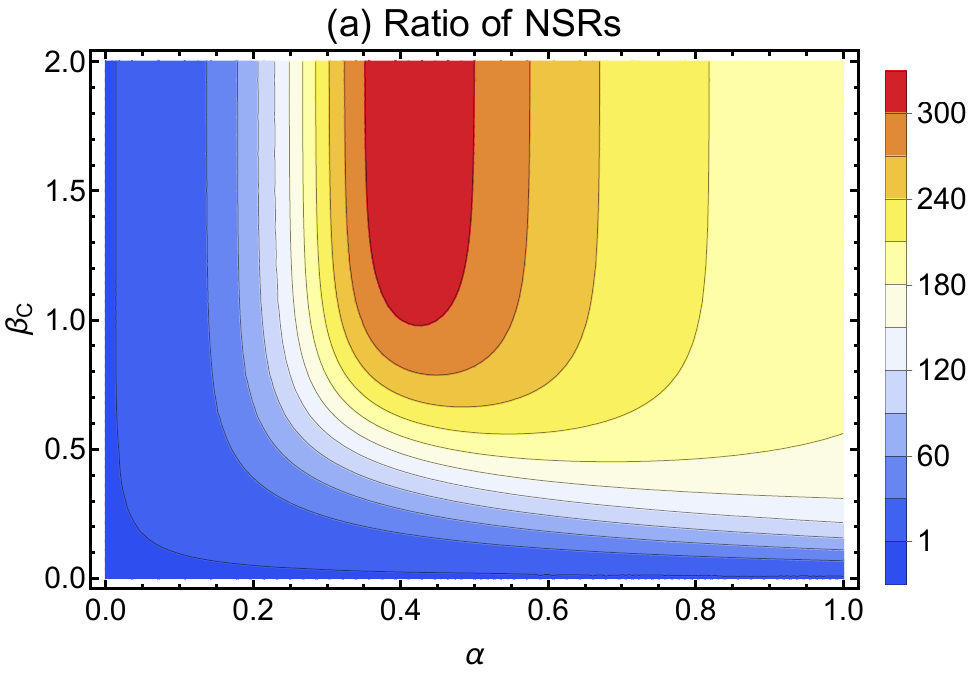}
\space
\space
\space
\includegraphics[width=5.3cm]{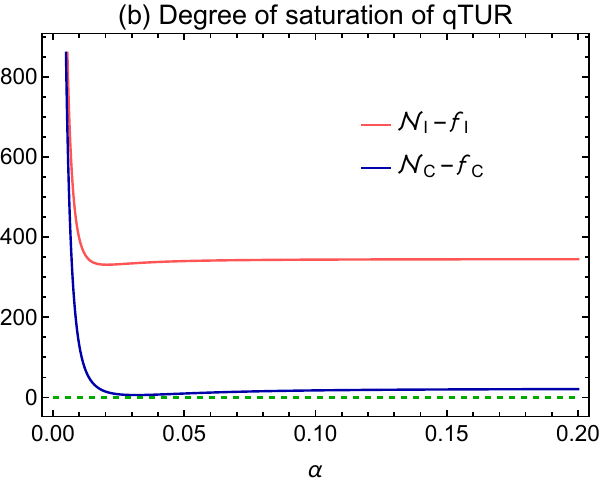}
\space
\space
\space
\includegraphics[width=5.3cm]{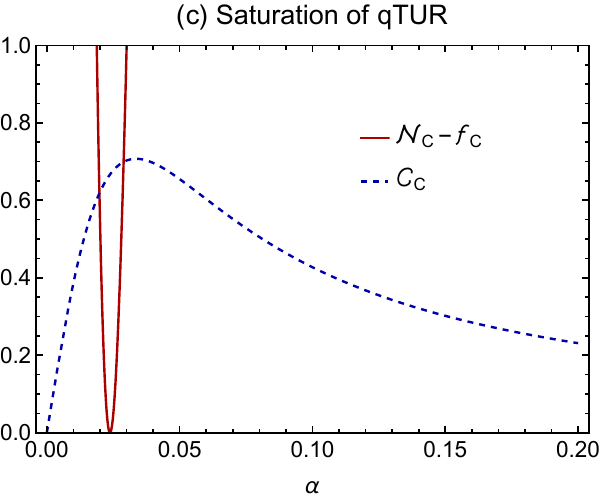}
 \caption{{\bf Comparisons of noise-to-signal ratios (NSRs) in coherent and incoherent engines.} The parameters $\gamma_{0} = 0.01$, $\omega_{h} = 10$, and $\omega_{c} = 5$ are considered for all the figures. (a) The figure on the left displays the ratio $\mathcal{N}_I/\mathcal{N}_C$ of NSRs in power (see Eq.~\eqref{eq:NSR}) corresponding to incoherent and coherent heat engines against $\beta_c$ and $\alpha$, while $\beta_h=0.001$. Note, $\mathcal{N}_I>\mathcal{N}_C $ signifies that the coherent engine produces less NSR in power than the incoherent engine, and the ratio can reach up to $\mathcal{N}_I/\mathcal{N}_C \geq 330$. (b) The figure in the middle shows the difference between the NSR and its lower bound for CQHEs and IQHEs (i.e., degree of saturation of qTUR) in Eq.~\eqref{eq:NSR}, involving NSRs and their quantum bounds with respect to $\alpha$ for $\beta_h=0.01$ and $\beta_c=0.8$. The traces in dark-blue and light-red represent $\mathcal{N}_C - f_C$ and $\mathcal{N}_I - f_I$ for the coherent and incoherent engines, respectively. The dashed-green trace corresponds to the zero value. (c) The figure on the right represents the saturation of qTUR by CQHEs for the parameters $\beta_h=0.01$ and $\beta_c=3$ with a large amount of energetic coherence. Here, $\mathcal{C}_C=\mathcal{C}(\sigma_C)$ represents the energetic coherence in the steady state of CQHEs. }
\label{fig:BLP}
\end{figure*}

{\it Power and efficiency} -- Now, we study power and efficiency. The power delivered by a steady state engine with resonant driving has a monotonic relation with the energetic coherence present in the steady state, and it is given by (see Appendices~\ref{appsec:IQHEsteaystate} and~\ref{appsec:CQHEsteaystate})
\begin{align}
\mathcal{P}_{X}=|\langle P_X \rangle|= \alpha \  (\omega_{h}-\omega_{c}) \ \mathcal{C}(\sigma_X), \label{eq:PtoC}    
\end{align}
which is a non-linear function of $\alpha$. As shown in Fig.~\ref{fig:CoherenceandPower}(a), it increases with $\alpha$. The power is proportional to coherence for a given $\alpha$. In fact, the ratio of the powers of CQHEs and IQHEs becomes equal to the ratio of the energetic coherence present in their respective steady states, i.e., $\mathcal{P}_{C}/\mathcal{P}_{I}=\mathcal{C}(\sigma_C)/\mathcal{C}(\sigma_I)$. Given that $\mathcal{C}(\sigma_C)>\mathcal{C}(\sigma_I)$ in general, the power of CQHEs is higher than the power delivered by IQHEs or $\mathcal{P}_{C}/\mathcal{P}_{I} >1$. A numerical analysis of the power ratio is presented in Fig.~\ref{fig:CoherenceandPower}(b) with respect to the bath temperatures and the driving strength, which displays that $ \mathcal{P}_{C}/\mathcal{P}_{I}$ is not only greater than 1, but can reach above 135, i.e. $\mathcal{P}_{C}/\mathcal{P}_{I} \geq 135$. Clearly, CQHEs exhibit quantum enhancements over IQHEs in power. The heat current from the hot bath is given by
\begin{align}
\langle \dot{J}_X^h \rangle= \alpha \  \omega_{h} \ \mathcal{C}(\sigma_X), \label{eq:JtoC}    
\end{align}
for both coherent and incoherent heat engines, and it has a monotonic relation with energetic coherence in the steady states. Yet again, due to energetic coherence, the heat current in CQHEs is higher than in IQHEs. In other words, the CQHEs have a higher capacity to draw heat from the hot bath than the IQHEs. However, the former also produces more power than the latter. Consequently, the efficiency $\eta_X=-\langle P_X \rangle/\langle \dot{J}_X^h \rangle$ remains same for both the engines, i.e.,
\begin{align}
\eta_I=\eta_C=1-\omega_c/\omega_h.    
\end{align}
Thus, CQHEs perform as good as IQHEs as far as heat-to-work conversion efficiency is concerned. See Appendices~\ref{appsec:IQHEsteaystate} and~\ref{appsec:FCS} for the derivations. \\

{\it Noise-to-signal ratio (NSR) of the power} -- In microscopic heat engines, power output often fluctuates. This, in turn, delimits the reliability or stability of the engines. The fluctuation is usually expressed in terms of the variance of power $\Delta P_X$, for $X=I, C$. For CQHEs and IQHEs, they are
\begin{align*}
\Delta P_X = \lambda_{1}^X|\langle P_{X} \rangle| - \lambda_{2}^X|\langle P_{X} \rangle|^{3},
\end{align*}
where coefficients $\lambda_{i}^X$s are functions of system and bath parameters. See Appendix~\ref{appsec:FCS} for more details.

Ideally, a good engine is expected to deliver high power output and low power output fluctuations. This quality is characterized by the NSR in power, i.e., the ratio between the fluctuation in power $\Delta P_X$, and the square of the average power output $\langle P_X \rangle^2$, and it is lower bounded by a quantum limit \cite{Hasegawa2020} as 
\begin{align}
\mathcal{N}_X = \frac{\rm \Delta P_X}{\langle P_X \rangle ^2} \geq f_X, \label{eq:NSR}   
\end{align}
where the lower bound $f_X$ is determined by quantum dynamical activity and coherent dynamical contribution. This relation is known as the quantum thermodynamic uncertainty relation (qTUR), and it is derived using quantum Cramér-Rao bound~\cite{Hasegawa2020}. The bounds $f_X$ in Eq.~\eqref{eq:NSR} are different for coherent and incoherent heat engines as they depend on the underlying Markovian dynamics (see Appendix~\ref{QTURR}). We find that the $\mathcal{N}_X$ depends on the energetic coherence present in the steady states and, for CQHEs and IQHEs with resonant driving, they are (see Appendix~\ref{appsec:FCS})
\begin{align}
    \mathcal{N}_{C}&= \frac{F_{p}}{\alpha \ \mathcal{C}(\sigma_C)}\left(1- \frac{3}{2}\mathcal{C}(\sigma_C)^2 \right), \label{eq:NC} \\
    \mathcal{N}_{I}&= \frac{F_{p}}{\alpha \ \mathcal{C}(\sigma_I)}\left(1- \frac{k}{F_{p}} \mathcal{C}(\sigma_I)^2\right), \label{eq:NI}
\end{align}
respectively, where 
\begin{align} 
F_{p}= \frac{2n_{h}n_{c}+n_{hc}}{n_{h}-n_{c}}, \ 
k =\frac{4\alpha^2 + \gamma_0^2(n_{hc}^2 + 2n_{hc} +3 n_{h}n_{c})}{\gamma_{0}^{2}(n_{h}-n_{c})}. \label{eq:Fp}
\end{align}
From the Eqs.~\eqref{eq:NC} and~\eqref{eq:NI}, it is seen that the NSR in both coherent and incoherent engines can be suppressed by accessing energetic coherence in the steady state for fixed $n_h$ and $n_c$. We observe that the NSR for CQHEs is, in general, much lower than that of IQHEs, which is the consequence of $\mathcal{C}(\sigma_C) \gg \mathcal{C}(\sigma_I)$. As shown in Fig.~\ref{fig:BLP}(a), the NSR in IQHEs can be more than 330 times of the NSR attained in CQHEs. Clearly, CQHEs are more reliable or deliver more precision in power than IQHEs.  

The saturation of the relation~\eqref{eq:NSR}, i.e., $\mathcal{N}_{X}=f_X$, implies that the engine is producing the least possible NSR in power that is given by its quantum bound. This is the best possible operating condition one would desire from an engine. A numerical analysis presented in Fig.~\ref{fig:BLP}(b) demonstrates that the CQHEs can operate in a regime where they yield very low NSR in power close to the quantum bound. In contrast, the IQHE has more NSR in power, which is far from its quantum bound. In addition, the CQHEs can saturate the qTUR by harnessing a large amount of energetic coherence, as shown in Fig.~\ref{fig:BLP}(c). Overall, the CQHEs are highly reliable and exhibit substantial quantum enhancements over IQHEs.\\

{\it Violations of cTUR} -- 
For classical heat engines, it is known that the rate of entropy production and the noise-to-signal ratio (NSR) in power follow a trade-off relation. This feature has been studied in terms of classical thermodynamic uncertainty relation (cTUR) ~\cite{Barato2015}, given by
\begin{align}
 \mathcal{Q}=\dot{S} \mathcal{N}  \geq 2, \label{eq:TUR}
\end{align}
where $\dot{S}=- \beta_{h} \langle \dot{J}_{h} \rangle - \beta_{c}\langle \dot{J}_{c} \rangle$ is the entropy production rate due to steady state dynamics and $\mathcal{N}=\Delta P/\langle P \rangle ^2$ is NSR in power. It implies that a reduction in NSR can be achieved at the cost of increasing the entropy production rate $\dot{S}$, particularly when the bound in~\eqref{eq:TUR} is saturated. 
This, in turn, represents more degree of irreversibility in the engine operation, leading to a reduced heat-to-work conversion efficiency. A similar conclusion is also drawn from another relation, known as the power-efficiency-constancy trade-off relation~\cite{Pietzonka2018}. Interestingly, it coincides with cTUR for CQHEs and IQHEs (see Appendix~\ref{CTUR}). 

We have discussed earlier that, for both coherent and incoherent heat engines, the NSR in power can be reduced while keeping the engine efficiency the same. This is why we witness violations of cTUR by CQHEs and IQHEs for some values of system-bath parameters, signifying that the engines can operate in the quantum regime. 

The left-hand side of relation~\eqref{eq:TUR} reduces to (for $X=I, C$) 
\begin{align}
\mathcal{Q}_X= \ln{\left(\frac{n_{h}(n_{c}+1)}{n_{c}(n_{h}+1)}\right)} F_X. \label{eq:QD}
\end{align}
Here $F_X=\frac{\Delta \dot{N}_X}{\langle\dot{N}_X\rangle}$ is the Fano factor, where ${\langle\dot{N}_X\rangle}={|\langle P_X\rangle}|/(\omega_h-\omega_c)$ is the photon current  and ${\Delta \dot{N}_X}=\Delta{P}_X/(\omega_{h}-\omega_{c})^2$ is the fluctuation in photon current. The violation of cTUR by CQHEs and IQHEs implies the violation of $\mathcal{Q}_C \geq 2$ and $\mathcal{Q}_I \geq 2$, respectively. Interestingly, the corresponding Fano factor can be expressed in terms of energetic coherence as 
\begin{align}
 F_{C}&= F_{p}\left(1- \frac{3}{2}\mathcal{C}(\sigma_C)^2\right), \label{eq:FanoC}  \\
 F_{I}&= F_{p}\left(1- \frac{k}{F_{p}} \mathcal{C}(\sigma_I)^2 \right), \label{eq:FanoI}
\end{align}
where $F_p$ and $k$ are given in Eq.~\eqref{eq:Fp} (see Appendix~\ref{CTUR}). In the absence of energetic coherence, $\mathcal{Q}=\ln{\left(\frac{n_{h}(n_{c}+1)}{n_{c}(n_{h}+1)}\right)}F_{p}$. In that case, the cTUR is respected because $\ln{\left(\frac{n_{h}(n_{c}+1)}{n_{c}(n_{h}+1)}\right)} F_p \geq 2$ \cite{Kalaee2021}. On the contrary, for quantum engines, the violations of cTUR can necessarily be attributed to the presence of energetic coherence in the steady states.

\begin{figure}
    \centering
    \includegraphics[width=7cm]{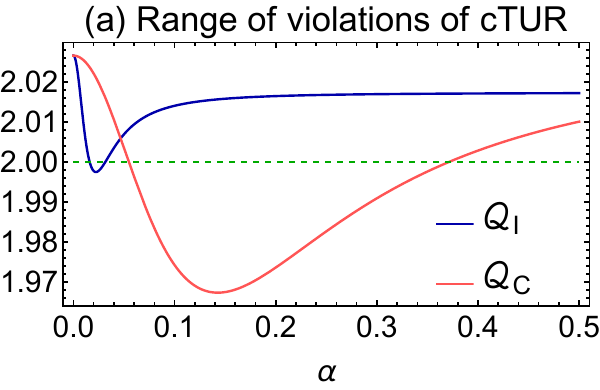}
\space 
\space 
\space \includegraphics[width=7cm]{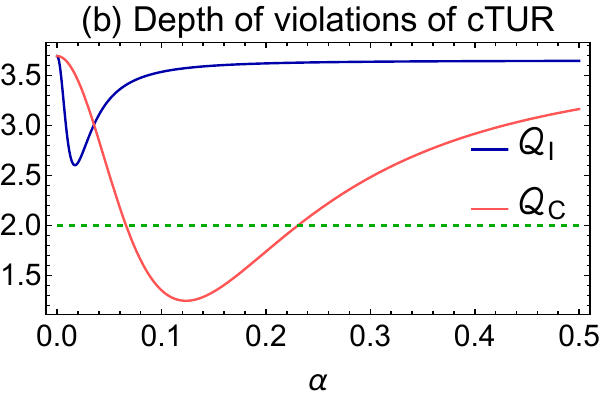}
 \caption{{\bf Violations of cTUR by CQHEs and IQHEs.} (a) The figure on the left displays the range of violations of cTUR by coherent and incoherent heat engines with respect to $\alpha$, for the parameters $\gamma_{0} = 0.01$, $\omega_{h} = 10$, $\omega_{c} = 5$, $\beta_h=0.01$, and $\beta_c=0.1$. (b) The figure on the right depicts the depth of violation of cTUR for the parameters $\gamma_{0} = 0.01$, $\omega_{h} = 10$, $\omega_{c} = 5$, $\beta_h=0.003$, and $\beta_c=0.7$. The figures show that the CQHE violates cTUR for a wider range of parameter $\alpha$. Further, the minimum value of $\mathcal{Q}_I$ is $1.997$ while the minimum value of $\mathcal{Q}_C$ can be $1.24$. See text for more details.}
    \label{fig:Q}
\end{figure}

The important point we highlight here is that the CQHEs violate cTUR not only for a wider range of parameters but also by a higher amount than IQHEs. This is, yet again, due to the fact that $\mathcal{C}(\sigma_C) > \mathcal{C}(\sigma_I)$ in general. A numerical study is carried out to compare $\mathcal{Q}_C$ and $\mathcal{Q}_I$ and presented in Fig.~\ref{fig:Q}(a) and~\ref{fig:Q}(b). We observe that $\mathcal{Q}_C$ can have values as low as $1.24$, while the lowest value of $\mathcal{Q}_I$ remains very close to $1.997$. Thus, IQHEs only marginally violate the classical limit. Overall, the violations of cTUR for a wider range of parameters and with a larger amount indicate that CQHEs possess more non-classical features than IQHEs. 

\section{Discussion \label{sec:Summary}}
The analysis and results presented above clearly demonstrate that, due to coherent heat transfers, coherent heat engines with resonant driving harness substantially higher energetic coherence in the working system than traditional (incoherent) quantum engines. Consequently, the power and noise-to-signal ratio in power is enhanced by hundreds of folds compared to their incoherent analogs. The noise-to-signal ratio in power has a fundamental lower bound derived from the quantum Cramér-Rao bound, and the inequality is termed the quantum thermodynamic uncertainty relation (qTUR). We have shown that coherent engines can yield a substantially low noise-to-signal ratio in power, which is very close to the lower bound (quantum limit). Even the CQHEs can saturate this quantum bound by harnessing high energetic coherence. This suggests that saturation of qTUR requires a high amount of coherence. Thus, coherent engines are highly reliable. In addition, unlike incoherent engines, coherent engines violate classical thermodynamic uncertainty relation for a much wider range of parameters and by a much higher amount. Altogether, the coherent engines possess more quantum features and greatly outperform conventional quantum and classical heat engines, manifesting genuine quantum enhancements.

 As we have discussed, for the resonant driving case, we can harness a high amount of energetic quantum coherence due to coherent heat transfer, which leads to an advantage in the power and reliability of coherent heat engines. However, in the context of standard SSD heat engines (which we refer to as incoherent heat engines) with non-resonant driving, as previously studied, the energetic coherence may lead to disadvantages in reliability, particularly when the detuning parameter is slightly larger~\cite{Kalaee2021}. We would like to emphasize that our study demonstrates the advantage of CQHEs over IQHEs due to coherent heat transfer rather than just the presence of energetic coherence. For the sake of completeness, we have analyzed the non-resonant case with arbitrary detuning, for both coherent and incoherent heat engines, and shown that coherent engines still outperform incoherent ones again due to coherent heat transfer in Appendix~\ref{NonReso}.

Two-photon Raman transitions provide a very common and standard tool in contemporary applications of quantum optics~\citep{Cohen1998, *Haroche2006, *Meystre2021, *Larson2021}. This paves the way for the realization of coherent quantum heat engines on various experimental platforms (see Appendix~\ref{exp}). Raman transitions have been easily demonstrated in various experimental setups, such as superconducting circuits~\cite{Novikov2016, Aamir2022, Zanner2022}, atom-optical systems~\cite{Zanon2005, Gauguet2008, Kristensen2021}, and nitrogen-vacancy centers in diamond~\cite{Bohm2021}, among others. Thus, our present analysis and results not only improve the understanding of quantum thermal devices, particularly how energetic coherence greatly enhances engine performance, but also open up new avenues for quantum-enabled technologies in the near future.

\section*{Acknowledgements}
R.G. thanks the Council of Scientific and Industrial Research (CSIR), Government of India, for financial support through a fellowship (File No. 09/947(0233)/2019-EMR-I). M.L.B. acknowledges financial support from the Spanish MCIN/AEI/10.13039/501100011033 grant PID2020-113334GB-I00, Generalitat Valenciana grant CIPROM/2022/66, the Ministry of Economic Affairs and Digital Transformation of the Spanish Government through the QUANTUM ENIA project call - QUANTUM SPAIN project, and by the European Union through the Recovery, Transformation and Resilience Plan - NextGenerationEU within the framework of the Digital Spain 2026 Agenda, and by the CSIC Interdisciplinary Thematic Platform (PTI+) on Quantum Technologies (PTI-QTEP+). This project has also received funding from the European Union’s Horizon 2020 research and innovation program under grant agreement CaLIGOLA MSCA-2021-SE-01-101086123. M.L. acknowledges financial supports from Europea Research Council AdG NOQIA; MCIN/AEI (PGC2018-0910.13039/501100011033, CEX2019-000910-S/10.13039/501100011033, Plan National FIDEUA PID2019-106901GB-I00, Plan National STAMEENA PID2022-139099NB, I00, project funded by MCIN/AEI/10.13039/501100011033 and by the ``European Union NextGenerationEU/PRTR'' (PRTR-C17.I1), FPI); QUANTERA MAQS PCI2019-111828-2); QUANTERA DYNAMITE PCI2022-132919, QuantERA II Programme co-funded by European Union's Horizon 2020 program under Grant Agreement No 101017733); Ministry for Digital Transformation and of Civil Service of the Spanish Government through the QUANTUM ENIA project call - Quantum Spain project, and by the European Union through the Recovery, Transformation and Resilience Plan - NextGenerationEU within the framework of the Digital Spain 2026 Agenda; Fundació Cellex; Fundació Mir-Puig; Generalitat de Catalunya (European Social Fund FEDER and CERCA program, AGAUR Grant No. 2021 SGR 01452, QuantumCAT \ U16-011424, co-funded by ERDF Operational Program of Catalonia 2014-2020); Barcelona Supercomputing Center MareNostrum (FI-2023-1-0013); EU Quantum Flagship PASQuanS2.1, 101113690, EU Horizon 2020 FET-OPEN OPTOlogic, Grant No 899794;  EU Horizon Europe Program (This project has received funding from the European Union’s Horizon Europe research and innovation program under grant agreement No 101080086 NeQSTGrant Agreement 101080086 — NeQST); European Union’s Horizon 2020 program under the Marie Sklodowska-Curie grant agreement No 847648; ICFO Internal “QuantumGaudi” project; ``La Caixa'' Junior Leaders fellowships, La Caixa” Foundation (ID 100010434): CF/BQ/PR23/11980043. Views and opinions expressed are, however, those of the author(s) only and do not necessarily reflect those of the European Union, European Commission, European Climate, Infrastructure and Environment Executive Agency (CINEA), or any other granting authority.  Neither the European Union nor any granting authority can be held responsible for them.

\onecolumngrid
\section*{Appendix}
Here, we include the derivations and analytical calculations to supplement the results presented in the main text.
\appendix

\section{Rotating frame and steady-state thermodynamics}\label{ROT}
The Lindblad dynamics of a driven system (where time dependence arises in the Hamiltonian due to the driving) with time-independent jump operators generally do not lead to a steady state. However, for a periodic time-dependence of driving Hamiltonian, as in \( H_d(t) \), there is a rotating frame in which the Hamiltonian can be made time-independent. For that, a counter-rotation is applied to the laboratory frame by $U = e^{i H_R t}$ with \([H_R, H_S] = 0\), where \( H_S \) is the internal Hamiltonian of the system, and $H_R$ is an arbitrary operator which commutes with $H_{S}$. In the rotating frame, an arbitrary operator \( B \) in the laboratory frame transforms as  
\[
B \to B_R = U B U^{\dagger}.
\]  
Further, there exists a Hamiltonian \( H_R \) for which the interaction Hamiltonian reduces to a time-independent one, given by  
\[
H_{dR} = U H_d(t) U^{\dagger}.
\]  
Accordingly, the overall transformed Hamiltonian becomes time-independent, and it is  
\[
{H}_{rot} = H_S - H_R + H_{dR}.
\]
The total Hamiltonian of CQHEs and IQHEs, i.e., $H_{S}(t) = H_{S} + H_d(t)$, where $H_d(t) = \alpha \left( e^{-i\omega_d t} \ketbra{1}{0} + e^{i\omega_d t} \ketbra{0}{1} \right)$, and for the choice $H_{R} = (\omega_{h} - \omega_{c} + \omega_d) \op{1} + \omega_h \op{2}$, the Hamiltonian in the rotating frame reduces to
\[
{H}_{rot} = -\delta \op{1} + H_{dR},
\]
where $H_{dR}=\alpha(\ketbra{1}{0}+\ketbra{0}{1})$ and detuning parameter $\delta=\omega_{d}-(\omega_{h}-\omega_{c})$. It is important to note that, for resonant driving, $\delta=0$ and  $\delta\neq0$ for non-resonant cases. In the rotating frame, for the steady state $\sigma_X$, with $\dot{\sigma}_X=0$ (see below), the average power $ \langle P_X \rangle $ of engines is given by
\begin{align}
\langle P_X \rangle=-i \  \Tr([H_S, H_{dR}] \sigma_C)= -\alpha \langle \dot{N}_{X} \rangle (\omega_h-\omega_c)  \leq 0,    \end{align}
where $\langle \dot{N}_{X} \rangle$ is the average photon flux. The photon flux is related to the imaginary part of the off-diagonal element of the density matrix, which is given as 
\begin{align}
\langle \dot{N}_{X} \rangle = 2\alpha \Im{\sigma^{X}_{ij}}. 
\end{align}
It is important to note that when the $\Re{\sigma^{X}_{ij}}=0$, then average photon flux is equal to $\alpha$ times the $l$-1 norm of coherence~\cite{Baumgratz2014}, i.e., $\langle \dot{N}_{X} \rangle = \alpha \mathcal{C}(\sigma_{X})$.

\section{ Incoherent Quantum Heat Engines (IQHEs)}\label{InCEng}
The interaction between the working system (qutrit) and baths is described by Hamiltonian~\eqref{eq:IntHamInCoh}. For weak system-baths coupling $|g_h|, |g_c| \ll 1$, the local dynamics of the qutrit is expressed by the Lindblad master equation \cite{Breuer2007, Boukobza2006b, Boukobza2006a, Boukobza2007}
\begin{align}
    \dot{\rho} = i \ [\rho, \ H_S(t)] + \mathcal{D}_h(\rho) + \mathcal{D}_c(\rho), \label{eq:IncMasEq}
\end{align}
where $\rho$ is the density matrix representing the state of the qutrit. The dissipators $\mathcal{D}_{h}(\rho)$ and $\mathcal{D}_{c}(\rho)$ represent dissipative dynamics due to the interactions with the hot and cold baths and are given by (for $x$ = $h$, $c$):
\begin{align*}
\mathcal{D}_{x}(\rho)=\gamma_{x}&(n_{x}+1)(b_{x} \rho b^{\dag}_{x} - \{b^{\dag}_{x}b_{x},\rho\}/2)\\ \nonumber 
&+ \gamma_{x}n_{x} (b^{\dag}_{x} \rho b_{x} - \{b_{x}b^{\dag}_{x},\rho\}/2),
\end{align*}
where the anti-commutator $\{Y,Z\} = YZ + ZY$, the coefficient $\gamma^x$ is the Weiskopf-Wigner decay constant, and $n_{x} = 1/(e^{\beta_{x}\omega_x}-1)$ is the average number of photons in the bath with frequency $\omega_x$. The appearance of two dissipators, $\mathcal{D}_{h}(\rho)$ and $\mathcal{D}_{c}(\rho)$, in the master equation \eqref{eq:IncMasEq} reflects that the heat exchange with the hot bath is independent (or uncorrelated) of the heat exchange with the cold baths. Thus, the heat exchanges between the baths are incoherent.

To quantify the power, heat currents, and other relevant quantities of IQHEs, we move to a rotating frame using a transformation $B_{R}=e^{iH_{R}t}Be^{-iH_{R}t}$, where $B$ is an arbitrary operator and $[H_{S},H_{R}]=0$ \cite{Boukobza2007, Bera2024}. This transformation eliminates the time dependence of $H_S(t)$ and reduces it to $H_{rot}=H_S - H_{R} + H_{dR}$, where $H_{dR}=\alpha(\ketbra{1}{0}+\ketbra{0}{1})$ (see Appendix~\ref{ROT}). The dissipators remain unchanged in the rotating frame, and the dynamics leads to a steady state $\sigma_I$ with $\dot{\sigma}_I=0$ (see Appendix~\ref{appsec:IQHEsteaystate}). Now the average power $\langle P_I \rangle$ and the average heat currents $\langle \dot{J}^{x}_I \rangle$ are given by 
\begin{align}
\langle P_I \rangle=-i\Tr([H_{S},H_{dR}]\sigma_I), \ \mbox{and} \ 
\langle \dot{J}^{x}_I \rangle = \Tr(\mathcal{D}_{x}(\sigma_I)H_{S}).
\end{align}
Note, $\langle P_I \rangle \leq 0$ for a heat engine, and the heat-to-work conversion efficiency is $\eta_I=-\langle P_I \rangle/\langle \dot{J}^{h}_I \rangle \geq 0$. Other important quantities, such as fluctuation in power ($\Delta P_I$) and fluctuation in heat currents ($\Delta J_I^x$), where power and heat currents are considered as random variables, are computed using full counting statistics of the steady state dynamics. See Appendix~\ref{appsec:FCS} for more details.

\section{Experimental feasibility of Coherent Quantum Heat Engines (CQHEs)}\label{exp}
The key difference between Coherent and Incoherent heat engines lies in the interaction between three-level systems (working systems) and hot and cold baths. In a coherent heat engine, the working system, hot bath, and cold bath interact via three-body interaction (see Eq.~\eqref{eq:IntHamCoh}), while in an incoherent heat engine system, interaction with the hot or cold bath is via two-body interaction independently (see Eq.~\eqref{eq:IntHamInCoh}). Both of these engine models can be realized with three-level atoms ($\Lambda$ type atoms) interacting with two external quantized electromagnetic fields at unequal temperatures. Important to note that these $\Lambda$ type atoms are extensively studied in the standard quantum optics literature for both cases when $\Lambda$ type atoms interact with two different electric fields independently via two-body interaction (one photon transition) or collectively via three-body interaction (for two-photon transition, i.e., referred as Raman transition), for details see Refs.~\citep{Gerry1990, Gerry1992, Wu1996,Cohen1998, *Haroche2006, *Meystre2021, *Larson2021}. Moreover, such a setup has also been experimentally realizable on various experimental platforms, such as atom-optical setup~\cite{Zanon2005} and superconducting circuits~\cite{Novikov2016}. Therefore, both engines are experimentally feasible and can be realized on the same experimental platform.

\section{Steady state solution of incoherent quantum heat engines in rotating frame \label{appsec:IQHEsteaystate}}
For incoherent engines, the total Hamiltonian of the qutrit system and two photonic (bosonic) thermal baths can be written as
\begin{equation}
    H = H_{S} + H_{B_{h}} + H_{B_{c}} + H_{SB_hB_c}^I,
\end{equation}
where the Hamiltonian and of the qutrit system is given by 
\begin{equation}
    H_{S}= \omega_{h} \ketbra{2} + (\omega_{h}-\omega_{c})\ketbra{1},
\end{equation}
with $\omega_{h}$ and  $\omega_{h}-\omega_{c}$ being the frequencies corresponding to the energy gaps. The Hamiltonians of the photonic thermal baths $H_{B_{h}}$ and $H_{B_{c}}$ and the interaction $H_{SB_hB_c}^I$ given in the main text. The corresponding Lindblad master equation describing the local dynamics of the qutrit is given in Eq.~(18) in the Methods. In a rotating frame, given by $B_{R}=e^{iH_{R}t}Be^{-iH_{R}t}$ and any operator $B$ and $[H_{0}, H_{R}]=0$, the master equation becomes
\begin{align}\label{IQHER}
    \dot{\rho_{R}} =& -i[H_{rot},\rho_{R}] + \mathcal{D}_{h}(\rho_{R}) + \mathcal{D}_{c}(\rho_{R}),
\end{align}
where $H_{rot}= -\delta \op{1} +\alpha(\ketbra{1}{0}+\ketbra{0}{1})$. For the resonant driving case, we consider $\delta =0$ (see Appendix~\ref{ROT}). Thus, the steady-state solution of the above master equation can be obtained by solving $\dot{\rho}_{R}=0$ (we denote the steady state by $\sigma_{I}$), and it is

\begin{align}
    \sigma_{I} &=  \frac{4 \alpha ^2 (\gamma_{c}+\gamma_{h}+\gamma_{c} n_{c}+\gamma_{h} n_{h})+\gamma_{c} \gamma_{h}n_{c} (n_{h}+1) (\gamma_{c} n_{c}+\gamma_{h} n_{h})}{4 \alpha ^2 (\gamma_{c} (3 n_{c}+2)+\gamma_{h} (3 n_{h}+2))+\gamma_{c} \gamma_{h} (3 n_{c} n_{h}+n_{c}+n_{h}) (\gamma_{c} n_{c}+\gamma_{h} n_{h})} \op{0} \nonumber\\
 &-  \frac{2 i \alpha  \gamma_{c} \gamma_{h} (n_{h}-n_{c})}{4 \alpha ^2 (\gamma_{c} (3 n_{c}+2)+\gamma_{h} (3 n_{h}+2))+\gamma_{c} \gamma_{h} (3 n_{c} n_{h}+n_{c}+n_{h}) (\gamma_{c} n_{c}+\gamma_{h} n_{h})}  \ketbra{0}{1} \nonumber\\ 
 &+\frac{2 i \alpha  \gamma_{c}\gamma_{h} (n_{h}-n_{c})}{4 \alpha ^2 (\gamma_{c} (3 n_{c}+2)+\gamma_{h} (3 n_{h}+2))+\gamma_{c} \gamma_{h} (3 n_{c} n_{h}+n_{c}+n_{h}) (\gamma_{c} n_{c}+\gamma_{h} n_{h})}  \ketbra{1}{0} \nonumber\\
&+  \frac{4 \alpha ^2 (\gamma_{c}+\gamma_{h}+\gamma_{c} n_{c}+\gamma_{h} n_{h})+\gamma_{c} \gamma_{h}n_{h} (n_{c}+1) (\gamma_{c} n_{c}+\gamma_{h} n_{h})}{4 \alpha ^2 (\gamma_{c} (3 n_{c}+2)+\gamma_{h} (3 n_{h}+2))+\gamma_{c} \gamma_{h} (3 n_{c} n_{h}+n_{c}+n_{h}) (\gamma_{c} n_{c}+\gamma_{h} n_{h})}\op{1} \nonumber  \\
&+  \frac{ (4 \alpha ^2+\gamma_{c} \gamma_{h}n_{c}n_{h}) (\gamma_{c} n_{c}+\gamma_{h} n_{h})}{4 \alpha ^2 (\gamma_{c} (3 n_{c}+2)+\gamma_{h} (3 n_{h}+2))+\gamma_{c} \gamma_{h} (3 n_{c} n_{h}+n_{c}+n_{h}) (\gamma_{c} n_{c}+\gamma_{h} n_{h})}\op{2} .
\end{align}

The $l$-1 norm of energetic coherence~\cite{Baumgratz2014} in the steady state on the rotating frame is  
\begin{align}
    \mathcal{C}(\sigma_{I}) = |\sigma_{I}^{(01)}| + |\sigma_{I}^{(10)}| = \frac{4 \alpha  \gamma_{c} \gamma_{h} (n_{h}-n_{c})}{4 \alpha ^2 (\gamma_{c} (3 n_{c}+2)+\gamma_{h} (3 n_{h}+2))+\gamma_{c} \gamma_{h} (3 n_{c} n_{h}+n_{c}+n_{h}) (\gamma_{c} n_{c}+\gamma_{h} n_{h})},
\end{align}
where $\sigma_{I}^{(ij)}=\bra{i} \sigma_{I} \ket{j}$. 
Now, the average power and the average heat currents in IQHEs corresponding to the hot and cold baths are given by
\begin{align}
&\langle P_{I} \rangle = -i\tr([H_{S},H_{dR}]\sigma_{I})=-i \alpha  (\omega_{h}-\omega_{c}) (\sigma^{(01)}_{I}-\sigma^{(10)}_{I}) =  - \alpha  (\omega_{h}-\omega_{c}) \ \mathcal{C}(\sigma_{I}), \\
&\langle \dot{J}^{h}_{I} \rangle= \Tr[\mathcal{D}_h(\sigma_I) H_S]= \alpha  \ \omega_h \ \mathcal{C}(\sigma_{I}), \\
 \mbox{and} \ \ & \langle \dot{J}^{c}_{I} \rangle= \Tr[\mathcal{D}_c(\sigma_I) H_S]= -\alpha  \ \omega_c \ \mathcal{C}(\sigma_{I}),
\end{align}
respectively. Accordingly, the heat-to-work conversion ratio for IQHEs is
\begin{equation}
    \eta_{I}= - \frac{\langle P_{I} \rangle }{\langle \dot{J}^{h}_{I} \rangle} = 1 -\frac{\omega_{c}}{\omega_{h}}.
\end{equation}

\section{Derivation of Lindblad master equation for coherent quantum heat engines \label{appsec:CQHEsteaystate}}
In this section, we derive the Lindblad master equation for a three-level quantum system coupled with the two photonic (bosonic) thermal baths (hot and cold baths), where the system and baths interact via two-photon transitions (Raman Interactions, i.e., three-body interactions). Our derivation follows the standard textbook approach discussed in Refs.~\cite{Breuer2007, Lidar2020}. The total Hamiltonian of the system and two photonic thermal baths can be written as

\begin{equation}\label{totalEq}
    H = H_{S} + H_{B_{h}} + H_{B_{c}} + H_{SB_hB_c}^C.
\end{equation}
where suffixes $h$ and $c$ correspond to hot and cold baths, respectively. We assume $\hbar = k_{B} = 1$ throughout this work. In Eq.~\eqref{totalEq}, the system Hamiltonian $H_S$ describes a three-level system (qutrit), given by
\begin{equation}
    H_{S}= \omega_{h} \ketbra{2} + (\omega_{h}-\omega_{c})b_{hc}^{\dagger}b_{hc} = \omega_{h} \ketbra{2} + (\omega_{h}-\omega_{c})\ketbra{1},
\end{equation}
where $\omega_{h}$ and  $\omega_{h}-\omega_{c}$ refers to the frequencies corresponding to the energy gaps, and $b_{hc}^{\dagger}=\ket{1}\bra{0}$ and $b_{hc}=\ket{0}\bra{1}$. In Eq.~\eqref{totalEq}, the photonic baths are a collection of infinite dimensional systems whose total Hamiltonian is given as
\begin{equation}
 H_{B_{h}} +   H_{B_{c}} = \sum_{k} \Omega_{k,h} a^{\dagger}_{k,h}a_{k,h} + \sum_{k'} \Omega_{k',c} a^{\dagger}_{k',c}a_{k',c}.
\end{equation}
Furthermore, in Eq.~\eqref{totalEq}, the interaction Hamiltonian between the system and the baths has the following form~\cite{Gerry1990, Gerry1992, Wu1996}
\begin{equation}
    H_{SB_hB_c}^C= g_0 \sum_{kk'}  (a_{k,h}a^{\dagger}_{k',c}b_{hc}^{\dagger}+a^{\dagger}_{k,h}a_{k',c}b_{hc}),
\end{equation}

Here, we consider system-baths coupling to be very weak, i.e.,  $g_0\ll 1$. The total Hamiltonian of the composite system (system + baths) in the interaction picture can be written as
\begin{equation}
   \Tilde{ H}(t) = g_{0} \sum_{k,k'} (a_{k,h}(t)a^{\dagger}_{k',c}(t)b_{hc}^{\dagger}(t) + a^{\dagger}_{k,h}(t)a_{k',c}(t)b_{hc}(t)),
\end{equation}
where $b_{hc}(t)=b_{hc}e^{ -i(\omega_{h}-\omega_{c})t}$, $b_{hc}^{\dagger}(t)=b_{hc}^{\dagger}e^{i(\omega_{h}-\omega_{c})t}$, $a_{p}(t)=a_{p}e^{-i\omega_{p} t}$ and $a_{p}^{\dagger}(t)=a_{p}^{\dagger}e^{i\omega_{p} t}$. For convenience, we can write the above Hamiltonian 
\begin{equation}
   \Tilde{ H}_{I}(t) = g_{0} \sum_{k,k'}  \sum_{\alpha=\{1,2\}} A_{\alpha}(t)\otimes B_{\alpha,kk'}(t),
\end{equation}
where $A_{1}(t)=b_{hc}^{\dagger}(t)$, $A_{2}(t)=b_{hc}(t)$, $B_{kk',1}(t)=a_{k,h}(t)a^{\dagger}_{k',c}(t)$ and $B_{kk',2}(t)=a^{\dagger}_{k,h}(t)a_{k',c}(t)$. In the interaction picture, the dynamics of the composite system is given by the von Neumann equation,
\begin{equation}
    \frac{d\Tilde{\rho}(t)}{dt}=-i[\Tilde{H}(t),\Tilde{\rho}(t)].
\end{equation}

For the cases where the system and baths are initially in a product state and very weakly coupled, using Born and Markov approximations, we obtain the following dynamical equation of the system
\begin{align}
    \frac{d\Tilde{\rho}(t)}{dt}=-g_{0}^2\sum_{\alpha \beta} \sum_{kk'ss'} \int_{0}^{\infty}d\tau\{\mathcal{B}_{\alpha \beta,kk'ss'}(
\tau,0)[A_{\alpha}(t),A_{\beta}(t-\tau)\Tilde{\rho}(t)]  
+\mathcal{B}_{ \beta \alpha,ss'kk'}(
0,\tau)[\Tilde{\rho}(t)A_{\beta}(t-\tau),A_{\alpha}(t)]\},
\end{align}
where $\mathcal{B}_{\alpha \beta,kk'ss'}(
\tau,0)=\tr(e^{iH_{B}\tau}B_{\alpha,kk'}e^{-iH_{B}\tau}B_{\beta,ss'}\rho_{\beta_{h}}\otimes \rho_{\beta_{c}} ) $ and $\mathcal{B}_{ \beta \alpha,ss'kk'}(0,
\tau)=\tr(B_{\beta,ss'}e^{iH_{B}\tau}B_{\alpha,kk'}e^{-iH_{B}\tau}\rho_{\beta_{h}}\otimes \rho_{\beta_{c}} )$. Here $H_{B}=H_{B_{h}}+H_{B_{c}}$ is total free Hamiltonian of the baths. The states $\rho_{\beta_{h}}$ and $\rho_{\beta_{c}}$ are the thermal states of hot and cold baths at inverse temperatures $\beta_{h}$ and $\beta_{c}$. The above dynamical equation in the frequency domain can be written as 
\begin{align}\label{feq}
\frac{d\Tilde{\rho}(t)}{dt}=-g_{0}^2 \sum_{kk'ss'}&\big[\int_{0}^{\infty}d\tau\mathcal{B}_{12,kk'ss'}(
\tau,0)e^{i\omega_{hc}\tau}[b_{hc}^{\dagger},b_{hc}\Tilde{\rho}(t)] 
+\int_{0}^{\infty}d\tau\mathcal{B}_{21,kk'ss'}(
\tau,0)e^{-i\omega_{hc}\tau}[b_{hc},b_{hc}^{\dagger}\Tilde{\rho}(t)] \nonumber\\ &+\int_{0}^{\infty}d\tau\mathcal{B}_{12,kk'ss'}(
0,\tau)e^{-i\omega_{hc}\tau}[\Tilde{\rho}(t)b_{hc}^{\dagger},b_{hc}]
+\int_{0}^{\infty}d\tau\mathcal{B}_{21,kk'ss'}(
0,\tau)e^{i\omega_{hc}\tau}[\Tilde{\rho}(t)b_{hc},b_{hc}^{\dagger}]  \big],
\end{align}
where $\omega_{hc}=\omega_{h}-\omega_{c}$. The bath correlation functions can be simplified as
\begin{align*}
    & \sum_{kk'ss'}\int_{0}^{\infty}d\tau\mathcal{B}_{12}(
\tau,0)e^{i\omega_{hc}\tau} =  \sum_{kk'ss'}\int_{0}^{\infty}d\tau \langle a_{k,h}(\tau)a^{\dagger}_{k',c}(\tau)a^{\dagger}_{s,h}a_{s',c}\rangle e^{i\omega_{hc}\tau} =  \sum_{kk'} (n_{k,h}(\Omega_{k,h})+1)n_{k',c}(\Omega_{k',c}) \int_{0}^{\infty}d\tau  e^{-i(\Delta_{kk',hc}-\omega_{hc})\tau}, \\
 & \sum_{kk'ss'} \int_{0}^{\infty}d\tau\mathcal{B}_{21}(
\tau,0)e^{-i\omega_{hc}\tau} = \sum_{kk'} (n_{k,c}(\Omega_{k,h})+1)n_{k',h}(\Omega_{k',c})\int_{0}^{\infty}d\tau  e^{i(\Delta_{kk',hc}-\omega_{hc})\tau},\\
 &\sum_{kk'ss'} \int_{0}^{\infty}d\tau\mathcal{B}_{12}(
0,\tau)e^{-i\omega_{hc}\tau} = \sum_{kk'} (n_{k,h}(\Omega_{k,h})+1)n_{k',c}(\Omega_{k',c})\int_{0}^{\infty}d\tau  e^{i(\Delta_{hc}-\omega_{kk',hc})\tau},\\
&\sum_{kk'ss'}\int_{0}^{\infty}d\tau\mathcal{B}_{21}(
\tau,0)e^{-i\omega_{hc}\tau} = \sum_{kk'} (n_{k,c}(\Omega_{k,h})+1)n_{k',c}(\Omega_{k',h})\int_{0}^{\infty}d\tau  e^{-i(\Delta_{kk',hc}-\omega_{hc})\tau}, \\
& \sum_{kk'ss'} \int_{0}^{\infty}d\tau  e^{\pm i(\Delta_{kk',hc}-\omega_{hc})\tau}= \sum_{kk'}   \pi \delta(\Delta_{kk',hc}-\omega_{hc}) \pm i \mathbb{P}(\frac{1}{\Delta_{kk',hc}-\omega_{hc}}),
\end{align*}
where $\Delta_{kk',hc}=\Omega_{k,h}-\Omega_{k',c}$. Here we have used the relations $\langle a^{\dagger}_{p}a_{p'}\rangle=n_{p}\delta_{pp'}$, $\langle a_{p}a^{\dagger}_{p'}\rangle=(n_{p}+1)\delta_{pp'}$ and $\langle a_{p}a_{p'}\rangle=\langle a^{\dagger}_{p}a^{\dagger}_{p}\rangle=0$ to simplify the bath correlation functions. To further simplify these functions, now we also convert $\sum _{p}\sum_{p'} =\int_{0}^{\infty} \int_{0}^{\infty}d\Omega d\Omega'  D(\Omega)D(\Omega')$, where $D(\Omega)$ is the photon density of states, i.e. the number of photon modes in a small frequency interval $[\Omega, \Omega + d \Omega ]$. Ignoring the principal value part for the moment, we then obtain 

\begin{align}
    \sum_{k,k'} f&\left(n_{h}(\Omega_{k,h}),n_{c}(\Omega_{k',c})\right) \int_{0}^{\infty}d\tau  e^{\pm i(\Delta_{kk',hc}-\omega_{hc})\tau} \nonumber \\
    &= \pi \int_{0}^{\infty}d\Omega_{h} D(\Omega_{h}) \int_{0}^{\infty}d\Omega_{c} D(\Omega_{c})  f\left(n_{h}(\Omega_{h}),n_{c}(\Omega_{c})\right) \delta\left((\Omega_{h} - \Omega_{c})-(\omega_{h} - \omega_{c})\right),
\end{align}
where $f$ is a function of $n_{h}(\Omega_{k,h})$ and $n_{c}(\Omega_{k',c})$. The double integral on the right-hand side is correlated. To match it with the incoherent quantum heat engines case, we enforce the resonance condition $( \Omega_{c}= \omega_{c})$ and $(\Omega_{h} =\omega_{h} )$. As a consequence, the expression reduces to
\begin{equation}
    \sum_{k,k'}  f(n_{h}(\Omega_{k,h}),n_{c}(\Omega_{k',c}))  \int_{0}^{\infty}d\tau  e^{\pm i(\Delta_{kk',hc}-\omega_{hc})\tau} = \pi \int_{0}^{\infty}d\Omega_{h} D(\Omega_{h}) \int_{0}^{\infty}d\Omega_{c} D(\Omega_{c})  f(n_{h}(\Omega_{h}),n_{c}(\Omega_{c}))  \delta(\Omega_{h} - \omega_{h} ) \delta(\Omega_{c}-\omega_{c}),
\end{equation}
and finally to
\begin{equation}
    \sum_{k,k'}  f(n_{h}(\Omega_{k,h}),n_{c}(\Omega_{k',c}))  \int_{0}^{\infty}d\tau  e^{\pm i(\Delta_{kk',hc}-\omega_{hc})\tau} = \pi   f(n_{h}(\omega_{h}),n_{c}(\omega_{c})) D(\omega_c) D(\omega_h).
\end{equation}

After substituting the expression of simplified bath correlation functions in Eq.~\eqref{feq}, we obtain the Lindblad master equation
\begin{align*}
    \frac{d\Tilde{\rho}(t)}{dt}=  \gamma_{1}\left(b_{hc}\Tilde{\rho}(t) b_{hc}^{\dagger} - \frac{1}{2}\{b_{hc}^{\dagger}     b_{hc},\Tilde{\rho}(t)\}\right)
    +  \gamma_{2}\left(b_{hc}^{\dagger}\Tilde{\rho}(t) b_{hc} - \frac{1}{2}\{b_{hc} b_{hc}^{\dagger},\Tilde{\rho}(t)\}\right),
\end{align*}
where $\gamma_1=\gamma_0n_c(n_h+1)$, $\gamma_2=\gamma_0 n_h(n_c+1)$, $\gamma_0=2g_{0}^{2} \pi D(\omega_c) D(\omega_h)$ is Weiskopf-Wigner decay constant, and $n_{x}=1/(e^{\beta_{x} \omega_{x}}-1)$ is average boson number of the bath '$x$' with inverse temperature $\beta_{x}$ ($x=h,c$). The Lindblad master equation derived above is in the interaction picture. It can be expressed in Schrodinger's Picture as
\begin{align*}
    \frac{d{\rho}(t)}{dt}&= -i [H_{S},\rho(t)] +  \gamma_{1}\left(b_{hc}{\rho}(t) b_{hc}^{\dagger} 
    - \frac{1}{2}\{b_{hc}^{\dagger}    b_{hc} ,{\rho}(t)\}\right) +  \gamma_{2}\left(b_{hc}^{\dagger}{\rho}(t) b_{hc} - \frac{1}{2}\{b_{hc}b_{hc}^{\dagger},{\rho}(t)\}\right).
\end{align*}
This dynamics leads to a steady state $\rho_{ss}$, i.e., $\dot{\rho}_{ss}=0$, given by 
\begin{equation}
    \rho_{ss} = \frac{\text{$\gamma $}_1 }{(\text{$\gamma $}_1+\text{$\gamma $}_2)}\op{0} + \frac{\text{$\gamma $}_2 }{(\text{$\gamma $}_1+\text{$\gamma $}_2)}\op{1}.
\end{equation}
For the steady state, the ratio of populations of exited state $\ket{1}$ and ground state $\ket{0}$ is given as 
\begin{equation}
    \frac{\rho^{(11)}_{ss}}{\rho^{(00)}_{ss}} = \frac{\gamma_2}{\gamma_1}= e^{{-(\beta_{h}\omega_{h}-\beta_{c}\omega_{c})}}=  e^{-\frac{(\beta_{h}\omega_{h}-\beta_{c}\omega_{c})}{(\omega_{h}-\omega_{c})}(\omega_{h}-\omega_{c})},
\end{equation}
where $\rho^{(ij)}_{ss}=\bra{i} \rho_{ss} \ket{j}$. To have an engine operation by utilizing two-photon transitions, we need population inversion, i.e., $\frac{\rho^{(11)}_{ss}}{\rho^{(00)}_{ss}} >1$. For this, the required condition is $\beta_{h}\omega_{h}-\beta_{c}\omega_{c} <0$. This also implies $n_h>n_c$.

\subsection*{Steady state solution of coherent quantum heat engines in rotating frame \label{appsec:CQHEpower}}
With an external periodic driving on the qutrit $H_d(t)= \alpha(e^{-i\omega_d t}\ketbra{1}{0}+e^{i\omega_d t}\ketbra{0}{1})$, the Lindblad master equation describing the dynamics of a coherent quantum heat engine becomes
\begin{align}
    \dot{\rho} =& -i[H_{S}+H_{d}(t),\rho]  +\mathcal{D}_{hc}(\rho),
\end{align}
where the master equation involves single dissipator $\mathcal{D}_{hc}(\rho)$, given by
\begin{align}
\mathcal{D}_{hc}(\rho)= \gamma_{1}(b_{hc}{\rho}(t) b_{hc}^{\dagger} 
    - \frac{1}{2}\{b_{hc}^{\dagger}    b_{hc} ,{\rho}(t)\}) +  \gamma_{2}(b_{hc}^{\dagger}{\rho}(t) b_{hc} - \frac{1}{2}\{b_{hc}b_{hc}^{\dagger},{\rho}(t)\}).   
\end{align}
We can transform the above Lindblad master equation to the rotating frame using the transformation $B_{R}=e^{iH_{R}t}Be^{-iH_{R}t}$, where $B$ is an arbitrary operator and $[H_{S}, H_{R}]=0$, as follows:
\begin{align}\label{CQHERot}
    \dot{\rho}_{R} =& -i[H_{rot},\rho_{R}] +\mathcal{D}_{hc}(\rho_{R}),
\end{align}

where $H_{rot}= -\delta \op{1} +\alpha(\ketbra{1}{0}+\ketbra{0}{1})$. For the resonant driving case, we consider $\delta =0$ (see Appendix~\ref{ROT}). 
 
A steady-state solution of the above master equation can be obtained by solving $\dot{\rho}_{R}=0$ (we denote the steady state by $\sigma_{C}$), which yields
\begin{equation}
    \sigma_{C} = \frac{4 \alpha ^2+\gamma_{1} (\gamma_{1}+\gamma_{2})}{8 \alpha ^2+(\gamma_{1}+\gamma_{2})^2}\op{0} + \frac{2 i \alpha  (\gamma_{1}-\gamma_{2})}{8 \alpha ^2+(\gamma_{1}+\gamma_{2})^2} \ketbra{0}{1} \\ 
    - \frac{2 i \alpha  (\gamma_{1}-\gamma_{2})}{8 \alpha ^2+(\gamma_{1}+\gamma_{2})^2}  \ketbra{1}{0} + \frac{4 \alpha ^2+\gamma_{2} (\gamma_{1}+\gamma_{2})}{8 \alpha ^2+(\gamma_{1}+\gamma_{2})^2}\op{1}.
\end{equation}
The $l$-1 norm of coherence~\cite{Baumgratz2014} of the steady state in CQHEs can be expressed as
\begin{align}
    \mathcal{C}(\sigma_{C}) = |\sigma_{C}^{(01)}| + |\sigma_{C}^{(10)}| = \frac{4  \gamma_{0}  \alpha (n_{h}-n_{c})}{8 \alpha ^2+ \gamma_{0}^2(2n_{h} n_{c} + n_{h} + n_{c})^2},
\end{align}
where $\sigma_{C}^{(ij)}=\bra{i} \sigma_{I} \ket{j}$. The average power is directly related to energetic coherence as
\begin{align}
&\langle P_{C} \rangle = -i\tr([H_{S},H_{dR}]\sigma_{C})=-i \alpha  (\omega_{h}-\omega_{c}) (\sigma^{(01)}_{C}-\sigma^{(10)}_{C}) =  - \alpha  (\omega_{h}-\omega_{c}) \ \mathcal{C}(\sigma_{C}).
\end{align}

The dynamics due to heat transfer with the baths is governed by single dissipator $\mathcal{D}_{hc}$, unlike in IQHEs discussed in Appendix~\ref{appsec:IQHEsteaystate}, and it takes into account the contributions from hot and cold baths together. Because of that, we cannot directly calculate the heat currents from the hot and cold baths with the dissipator. We overcome this limitation by employing the full counting statistics (FCS) of the steady-state dynamics in the rotating frame (see Appendix~\ref{appsec:FCS}). 

\section{Comparison of energetic coherences in coherent and incoherent heat engines \label{appsec:AlphaCrit}}
\begin{figure}
    \centering
 \includegraphics[width=10cm]{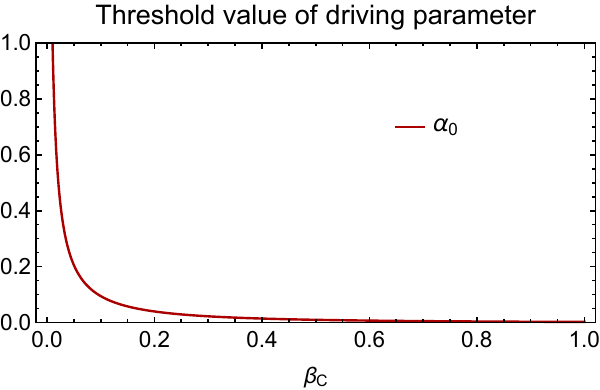}
 \caption{ The figure depicts the variation of the threshold value of the driving parameter ($\alpha_0$) against the inverse temperature of the cold bath. Here we consider $\gamma_0$ = 0.01, $\omega_h$ = 10, $\omega_c$ = 5, and $\beta_h$ = 0.001.  \label{appfig:AlphaCrit}}
\end{figure}

The energetic coherence in the steady state of the qutrit is non-linearly dependent on the driving parameter $\alpha$ for both coherent and incoherent heat engines. It is, in general, higher in the coherent heat engines compared to the incoherent ones. However, for some values of $\alpha$, the energetic coherence in the coherent heat engines can be lower than the incoherent ones. The driving parameter has a threshold value, given by $\alpha_{0}$, below which the energetic coherence is higher for incoherent heat engines. We determine the $\alpha_{0}$ by solving the condition
\begin{equation}
    \mathcal{C}(\sigma_C)=\mathcal{C}(\sigma_I),
\end{equation}
and it is
\begin{align}
\alpha_{0}  = \gamma_{0}  \sqrt{\frac{(n_{h}+n_{c}) (n_{h}+n_{c}+3 n_{h} n_{c})-\gamma_{0} ^2 (n_{c}+n_{h}+2n_{c}n_{h})^2}{8 -4  \left(3( n_{h}+n_{c})+4\right)}}.
\end{align}
The $\mathcal{C}(\sigma_C)>\mathcal{C}(\sigma_I)$ for $\alpha > \alpha_{0}$ and $\mathcal{C}(\sigma_C) \leq\mathcal{C}(\sigma_I)$ for $\alpha \leq \alpha_{0}$. Note that we need to satisfy the condition $n_h > n_c$ for the continuous device to operate as a heat engine. However, for reasonable values of the parameters $n_h$, $n_c$, and $\gamma_0$, the threshold value $\alpha_{0}$ remains very small, corresponding to a very weak external driving. Fig.~\ref{appfig:AlphaCrit} illustrates how $\alpha_{0}$ varies with respect to inverse temperatures of the baths. In the exceptional cases where the baths are extremely hot, i.e., $n_h \approx n_c \gg 1$, the $\alpha_{0}$ becomes very high. Nevertheless, considering the usual experimental situations, the engines operate with $\alpha > \alpha_{0}$, and the coherent engines yield more energetic coherence in their steady state than the incoherent engines.

\section{Full Counting Statistics \label{appsec:FCS}}
Full Counting Statistics (FCS) provides an analytical approach to determine the statistics of the quantity of interests $M$, such as power, currents corresponding to each bath, and their fluctuations in an open quantum system dynamics~\cite{Esposito2009}. This approach incorporates counting fields into the master equation. Suppose $\rho(\chi,t)$ represents the solution of the dressed Lindblad master equation. In that case, we define the moment-generating function $\mathcal{M}(\chi,t)$ and the cumulant-generating function $\mathcal{F}(\chi,t)$ as follows:
\begin{align}
\mathcal{M}(\chi,t)=\mathrm{tr}{\{\rho(\chi,t)\}}, \ \mbox{and} \ \mathcal{F}(\chi,t)=\ln \mathcal{M}(\chi,t).
\end{align}
Sometimes, a description in terms of cumulants is more convenient. The advantage lies in the fact that the dominant eigenvalue of the Liouvillian usually determines the long-time evolution of the cumulant-generating function:
\begin{equation}
 \mathcal{C}(\chi,t) \approx \lambda(\chi)t,
\end{equation}
where $\lambda(\chi)$is the eigenvalue of $\mathcal{L}(\chi)=\mathcal{L}(\chi,0)$ with the largest real part (uniqueness assumed) and it vanishes when $\chi = 0$.

In the long-time limit, the cumulants of the quantity of interest $M$ in the steady state can be obtained using the following formula:
\begin{equation}
\langle \langle M^{k} \rangle \rangle = \bigg(\frac{d}{d(i\chi)}\bigg)^{k} \lambda(\chi)\bigg|_{\chi=0}.
\end{equation}
The first and second cumulants correspond to the mean and variance of the quantity of interest $M$, respectively:
\begin{align}
\langle M \rangle = \bigg(\frac{d}{d(i\chi)}\bigg) \lambda(\chi)\bigg|_{\chi=0}, \ \ \mbox{and} \ \ \Delta M = \langle \langle M^{2} \rangle \rangle = \bigg(\frac{d}{d(i\chi)}\bigg)^{2} \lambda(\chi)\bigg|_{\chi=0}.
\end{align}

A direct computation of $\lambda(\chi)$ is not straightforward. To analytically determine the mean and variance from the derivatives, we follow the method outlined in Refs.~\cite{Bruderer2014, Kalaee2021, Prech2023, Landi2024}. Consider the characteristic polynomial of $\cal{L}(\chi)$
\begin{equation}
    \sum_{n}a_{n} \lambda(\chi)^{n} =0,
\end{equation}
where the terms $a_n$ are functions of $\chi$. Derivatives of $a_n$ are defined as 
\begin{align}
    a'_{n} = i\frac{d}{d\chi} a_{n}|_{\chi=0},  \ \mbox{and} \  a''_{n} = \big(i\frac{d}{d\chi}\big)^{n} a_{n}|_{\chi=0}.
\end{align}
With a little analysis, we can express mean and variance as (for more details, see Appendices of Refs.~\cite{Kalaee2021, Prech2023, Landi2024}):
\begin{align}
\langle M \rangle =-\frac{a'_{0}}{a_{1}},  \ \mbox{and} \ \Delta M = \bigg(\frac{a''_0}{a'_0}-\frac{2a'_1}{a_1}\bigg)\langle M \rangle -\frac{2a_2}{a_1} \langle M \rangle^2.
\end{align}
Note that the above expressions of mean and variance hold for all systems with Lindblad dynamics with a unique steady state.

\subsection{Counting field statistics for coherent quantum heat engines \label{appsec:FCSCohEng}}
Here, we re-derive the Lindblad master equation of coherent heat engine by introducing counting fields, which will help us to evaluate current and power statistics~\cite{Esposito2009}. The total Hamiltonian for the system and the baths (in the presence of driving) is given as
\begin{equation}
    H = H_{S}+H_{d}(t) + H_{B_{h}} + H_{B_{c}} + H_{SB_{h}B_{c}}^C.
\end{equation}
where $H_{d}(t)$ represents the external driving field acting on the three-level system and the rest of the Hamiltonians defined in the previous section. Here, we are considering a situation where the two baths continuously interact with the system, and the interaction between the system and the baths is weak. We choose the initial state as a product state, i.e., $\rho(0)=\rho_S(0)\otimes \rho_B$, where $\rho_B=\rho_{\beta_{h}}\otimes \rho_{\beta_{c}}$ and the baths are prepared in thermal states with respective Hamiltonians $H_{B_h}$, $H_{B_c}$ and inverse temperatures $\beta_h$ and $\beta_c$, respectively.  To measure the observables $H_{B_h}$ and $H_{B_c}$ and to get the corresponding probability distributions of their measurement, we introduce counting field $\chi_j$ $(j=h, c)$ to each bath. We introduce $\chi\equiv{\{\chi_h,\chi_c\}}$ to denote collectively both the counting variables. The modified density matrix $\rho(\chi,t)$ of composite system is given as
\begin{equation}
\label{EOM1}
\rho(\chi,t)=U(\chi,t)\rho(0)\bar{U}(-\chi,t),
\end{equation}
with
\begin{align*}
U(\chi,t)=e^{-i(\chi_h H_{B_h}+\chi_c H_{B_c})/2}U(t)e^{i(\chi_h H_{B_h}+\chi_c H_{B_c})/2}  \ \ \mbox{and} \ \  \bar{U}(-\chi,t)=e^{i(\chi_h H_{B_h}+\chi_c H_{B_c})/2}U^{\dagger}(t)e^{-i(\chi_h H_{B_h}+\chi_c H_{B_c})/2}
\end{align*}
being the counting field-dressed evolution operator. Here $U(t)$ is the unitary evolution operator generated by the total Hamiltonian $H$. The time evolution of modified density matrix $\rho(\chi,t)$ is given by following master equation
\begin{equation}
\frac{d\rho(\chi,t)}{dt}=-i[H(\chi,t)\rho(\chi,t)-\rho(\chi,t)H(-\chi,t)],
\label{rhoeq1}
\end{equation}
where, $H(\chi,t)=e^{-i(\chi_h H_{B_h}+\chi_c H_{B_c})/2}He^{i(\chi_h H_{B_h}+\chi_c H_{B_c})/2}$.
In the interaction picture, one gets (the operators are labeled by tilde)
\begin{equation}
\tilde{\rho}(\chi,t)=U_0\rho(\chi,t)U^{\dagger}_0, 
\end{equation}
where $U_0$ is the unitary operator generated by the Hamiltonian $H_0(t)=H_S(t)+H_B$. Here we have denoted $H_{S}(t)=H_{S}+H_{d}(t)$ and $H_{B}=H_{B_h}+H_{B_c}$. In the interaction picture, the dressed total Hamiltonian is given by
\begin{equation}
\tilde{H}_I(\chi,t)=U_0 H_{SB}(\chi,t)U_0^\dagger=\sum_{\alpha,kk'}A_{\alpha}(t)\otimes B_{\alpha,kk'}(\chi,t) \ \ \mbox{and} \ \ \tilde{H}_I(-\chi,t)=U_0 H_{SB}(\chi,t)U_0^\dagger=\sum_{\alpha,kk'}A_{\alpha}(t)\otimes B_{\alpha,kk'}(-\chi,t),
\end{equation} 
where $B_{\alpha,kk'}(\chi,t) = B^{h}_{\alpha,k}(\chi_h,t) \otimes B^{c}_{\alpha,k'}(\chi_c,t)$ and  $B^{h(c)}$'s are bath operators corresponding to hot (cold) bath. In the interaction picture, the evolution equation can be written as
\begin{equation}
\frac{d\tilde{\rho}(\chi,t)}{dt}=-i[\tilde{H}_I(\chi,t)\tilde{\rho}(\chi,t)-\tilde{\rho}(\chi,t)\tilde{H}_I(-\chi,t)].
\end{equation}
Next, considering the weak coupling assumption and performing the standard Born-Markov approximation, we arrive at the following master equation
\begin{align}
\frac{d\tilde{\rho}_S(\chi,t)}{dt}=-\int_0^{\infty}d\tau  {\rm Tr}_B[\tilde{H}_I & (\chi,t)\tilde{H}_I(\chi,t-\tau)\tilde{\rho}_S(\chi,t)\rho_B
-\tilde{H}_I(\chi,t) \tilde{\rho}_S(\chi,t) \rho_B\tilde{H}_I(-\chi,t-\tau)\nonumber\\
&-\tilde{H}_I(\chi,t-\tau)\tilde{\rho}_S(\chi,t)\rho_B\tilde{H}_I(-\chi,t)
+\tilde{\rho}_S(\chi,t)\rho_B\tilde{H}_I(-\chi,t-\tau)\tilde{H}_I(-\chi,t)],
\end{align}
where we have used ${\rm Tr}_B[\tilde{H}_I(\chi,t) \ \rho_B]=0$, and $\rho_{B}=\rho_{\beta_h}\otimes \rho_{\beta_c}$. After simplification, the above equation can be written as
\begin{align*}
\frac{d\tilde{\rho}_S(\chi,t)}{dt}=-g_{0}^2\int_0^{\infty}d\tau\sum_{\alpha \beta kk'ss'}&\left({A}_\alpha(t){A}_\beta(t-\tau)\tilde{\rho}_S(\chi,t)\Tr[{B}_{\alpha,kk'}(\chi,t){B}_{\beta,ss'}(\chi,t-\tau)\rho_B] \right. \\ 
& -{A}_\alpha(t)\tilde{\rho}_S(\chi,t){A}_\beta(t-\tau)\Tr[{B}_{\alpha,kk'}(\chi,t)\rho_B{B}_{\beta,ss'}(-\chi,t-\tau)]\\
\nonumber
&-{A}_\alpha(t-\tau)\tilde{\rho}_S(\chi,t){A}_\beta(t)\Tr[{B}_{\alpha,kk'}(\chi,t-\tau)\rho_B{B}_{\beta,ss'}(-\chi,t)] \\ & \left. +\tilde{\rho}_S(\chi,t){A}_\alpha(t-\tau){A}_\beta(t)\Tr[\rho_B{B}_{\alpha,kk'}(-\chi,t-\tau){B}_{\beta,ss'}(-\chi,t)]\right).
\end{align*}

After further simplifying the bath correlation function, we obtain
\begin{align}
\nonumber
\frac{d\tilde{\rho}_S(\chi,t)}{dt}=&-g_{0}^2\int_0^{\infty}d\tau\sum_{\alpha \beta}\sum_{kk'ss'}({A}_\alpha(t){A}_\beta(t-\tau)\tilde{\rho}_S(\chi,t)\tr[{B}_{\alpha,kk'}(\tau){B}_{\beta,ss'}(0)\rho_B] -{A}_\alpha(t)\tilde{\rho}_S(\chi,t){A}_\beta(t-\tau)\tr[{B}_{\beta,ss'}(-2\chi,\tau){B}_{\alpha,kk'}(0)\rho_B]\\
\nonumber
&-{A}_\alpha(t-\tau)\tilde{\rho}_S(\chi,t){A}_{\beta}(t)\tr[{B}_{\beta,ss'}(-2\chi,t){B}_{\alpha,kk'}(0)\rho_B]
+\tilde{\rho}_S(\chi,t){A}_\alpha(t-\tau){A}_\beta(t)\tr[{B}_{\alpha,kk'}(-\tau){B}_{\beta,ss'}(0)\rho_B]).
\end{align}
Using the explicit form of system and bath operators $A_{1}(t)=b_{hc}^{\dagger}(t)$, $A_{2}(t)=b_{hc}(t)$, $B_{kk',1}(t)=a_{k,h}(t)a^{\dagger}_{k',c}(t)$, $B_{kk',2}(t)=a^{\dagger}_{k,h}(t)a_{k',c}(t)$, $b_{hc}(t)=b_{hc}e^{ -i(\omega_{h}-\omega_{c})t}$, $b_{hc}^{\dagger}(t)=b_{hc}^{\dagger}e^{i(\omega_{h}-\omega_{c})t}$, $a_{p}(t)=a_{p}e^{-i\omega_{p} t}$ and $a_{p}^{\dagger}(t)=a_{p}^{\dagger}e^{i\omega_{p} t}$, and solving the bath correlation function and converting sums into integrals as considered in the previous section~\ref{appsec:CQHEsteaystate}, we get the following dressed Lindblad master equation in Schrodinger picture as
\begin{align}
     \frac{d{\rho}(\chi,t)}{dt} = -i[H_{S}+H_{d}(t),\rho] +  \gamma_{1}(e^{i(\omega_h\chi_h-\omega_c\chi_c)}b_{hc}{\rho}(t) b_{hc}^{\dagger} 
    - \frac{1}{2}\{b_{hc}^{\dagger}    b_{hc} ,{\rho}(t)\}) +  \gamma_{2}(e^{-i(\omega_h\chi_h-\omega_c\chi_c)}b_{hc}^{\dagger}{\rho}(t) b_{hc} - \frac{1}{2}\{b_{hc}b_{hc}^{\dagger},{\rho}(t)\}),
\end{align}

where $\gamma_1=\gamma_0n_c(n_h+1)$, $\gamma_2=\gamma_0 n_h(n_c+1)$, $\gamma_0=2g^{2} \pi D(\omega_c) D(\omega_h)$ is Weiskopf-Wigner decay constant, and $n_{x}=1/(e^{\beta_{x} \omega_{x}}-1)$ is the average photon number in the bath with inverse temperature $\beta_{x}$. In the rotating frame, the above master equation reduces to 
\begin{align}
     \frac{d{\rho}(\chi,t)}{dt} = -i[H_{rot},\rho] +  \gamma_{1}(e^{i(\omega_h\chi_h-\omega_c\chi_c)}b_{hc}{\rho}(t) b_{hc}^{\dagger} 
    - \frac{1}{2}\{b_{hc}^{\dagger}    b_{hc} ,{\rho}(t)\}) +  \gamma_{2}(e^{-i(\omega_h\chi_h-\omega_c\chi_c)}b_{hc}^{\dagger}{\rho}(t) b_{hc} - \frac{1}{2}\{b_{hc}b_{hc}^{\dagger},{\rho}(t)\}),
\end{align}
and the corresponding full Liouvillian super-operator with counting fields is
\begin{equation}
    \mathcal{L}(\chi_h,\chi_c)= \left(
\begin{array}{cccc}
 -\text{$\gamma $}_1 & -i \alpha  & i \alpha  & \text{$\gamma $}_2 e^{-i   (\chi_h\text{$\omega $}_h-\chi_c\text{$\omega $}_c)} \\
 -i \alpha  & -\frac{\text{$\gamma $}_1}{2}-\frac{\text{$\gamma $}_2}{2} & 0 & i \alpha  \\
 i \alpha  & 0 & -\frac{\gamma_1}{2}-\frac{\text{$\gamma $}_2}{2} & -i \alpha  \\
 \text{$\gamma $}_1 e^{i  (\chi_h\text{$\omega $}_h-\chi_c\text{$\omega $}_c)} & i \alpha  & -i \alpha  & -\text{$\gamma $}_2 \\
\end{array}
\right).
\end{equation}
where $H_{rot}= -\delta \op{1} +\alpha(\ketbra{1}{0}+\ketbra{0}{1})$ and for the resonant driving case, we consider $\delta =0$ (see Appendix~\ref{ROT}). For calculating power statistics, we set $\chi_{h}=\chi_{c}=\chi$. Following the previous discussion in this section, we can determine the polynomial factors with respective derivatives
\begin{align*}
    a_1&=2 \alpha ^2 ({\gamma }_1+{\gamma}_2)+\frac{1}{4} ({\gamma}_1+{\gamma}_2)^3,\\
    a_2&=\frac{1}{4} \left(16 \alpha ^2+5 ({\gamma}_1+{\gamma}_2)^2\right),\\
    a'_{0}&=\alpha ^2 ({\gamma }_1-{\gamma_2}) ({\gamma }_1+{\gamma }_2) ({\omega }_h-{\omega }_c),\\
    a''_{0}&=-\alpha ^2 ({\gamma}_1+{\gamma }_2)^2 ({\omega }_h-{\omega }_c)^2,\\
   \mbox{and} \ \ \  a'_1&=2 \alpha ^2 ({\gamma }_1-{\gamma }_2) ({\omega }_h-{\omega }_c).
\end{align*}
The expression for the average (mean) and variance of power are given by
\begin{align}
\langle P_{C} \rangle =\frac{4 \alpha ^2 (\text{$\gamma $}_1-\text{$\gamma $}_2) }{8 \alpha ^2+(\text{$\gamma $}
_1+\text{$\gamma $}_2)^2}(\text{$\omega $}_h-\text{$\omega $}_c),
\ \mbox{and} \ 
\Delta P_{C}=F_{p}(|\langle P_{C}\rangle| - \frac{3}{2\alpha^2(\omega_h-\omega_c)^2}|\langle P_{C}\rangle|^3  )(\omega_h-\omega_c),
\end{align}
where $F_{p}= \frac{2n_{h}n_{c}+n_{h}+n_{c}}{n_{h}-n_{c}}$. Similarly, we can determine the average and variance of heat current corresponding to a bath with inverse temperature $\beta_x$ by setting $\chi_{x}=\chi$ and $\chi_{y}=0$ in the Liouvillian super-operator. The average heat currents from the hot and cold baths are 
\begin{align}
    \langle \dot{J}^{h}_{C} \rangle =\frac{4 \alpha ^2 (\text{$\gamma $}_2-\text{$\gamma $}_1) }{8 \alpha ^2+(\text{$\gamma $}
_1+\text{$\gamma $}_2)^2}\text{$\omega $}_h, \ \ \mbox{and} \ \ \langle \dot{J}^{c}_{C} \rangle = \frac{4 \alpha ^2 (\text{$\gamma $}_1-\text{$\gamma $}_2) }{8 \alpha ^2+(\text{$\gamma $}
_1+\text{$\gamma $}_2)^2}\text{$\omega $}_c,
\end{align}
respectively, and the corresponding variances in heat currents are
\begin{align}
\Delta \dot{J}^{h}_{C}&=\frac{4 \alpha ^2 (\text{$\gamma $}_1+\text{$\gamma $}_2)  \left(64 \alpha ^4-8 \alpha ^2 \left(\text{$\gamma $}_1^2-10 \text{$\gamma $}_1 \text{$\gamma $}_2+\text{$\gamma $}_2^2\right)+(\text{$\gamma $}_1+\text{$\gamma $}_2)^4\right)}{\left(8 \alpha ^2+(\text{$\gamma $}_1+\text{$\gamma $}_2)^2\right)^3}\text{$\omega $}_h^2, \\
\mbox{and} \ \ \Delta   \dot{J}^{c}_{C}&=\frac{4 \alpha ^2 (\text{$\gamma $}_1+\text{$\gamma $}_2)  \left(64 \alpha ^4-8 \alpha ^2 \left(\text{$\gamma $}_1^2-10 \text{$\gamma $}_1 \text{$\gamma $}_2+\text{$\gamma $}_2^2\right)+(\text{$\gamma $}_1+\text{$\gamma $}_2)^4\right)}{\left(8 \alpha ^2+(\text{$\gamma $}_1+\text{$\gamma $}_2)^2\right)^3}\text{$\omega $}_c^2.
\end{align}
With this, the heat-to-work conversion efficiency of CQHEs becomes
\begin{equation}
    \eta_{C}= - \frac{\langle P_{C}\rangle}{\langle \dot{J}^{h}_{C}\rangle} = 1 -\frac{\omega_{c}}{\omega_{h}}.
\end{equation}
It is important to note that IQHEs and CQHEs have the same efficiency. Further, the noise-to-signal ratio of the power of CQHEs is
\begin{equation}
 \mathcal{N}_{C}=  \frac{ \Delta P_{C}}{\langle P_{C} \rangle^2}=F_{p}\big(\frac{1}{|\langle P_{C}\rangle| }- \frac{3}{2\alpha^2(\omega_h-\omega_c)^2}|\langle P_{C}\rangle|  \big)(\omega_h-\omega_c)=\frac{F_{p}}{\alpha \mathcal{C}(\sigma_{{C}})}(1- \frac{3}{2} {\mathcal{C}(\sigma_{{C}})^2}),
\end{equation}
where $\langle P_{C} \rangle =- \alpha (\omega_{h} - \omega_{c}) \mathcal{C}(\sigma_{C})$, and $\mathcal{C}(\sigma_{{C}})$ is $l$-1 norm of coherence of the steady state $\sigma_C$. It is important to note that the noise-to-signal ratio of currents, power, and photon number flux is the same for CQHEs.

\subsection{Counting field statistics for Incoherent quantum heat engines \label{appsec:FCSIncohEng}}
To determine the power statistics in incoherent heat engines, we again use the Full Counting Statistics (FCS) technique, which includes counting fields in the master equation. Let $\chi_{h}$ and $\chi_{c}$ be counting fields for the hot and cold baths, respectively. The dressed Lindblad master equation~\eqref{IQHER} of IQHEs in the rotating frame becomes
\begin{align}
    \dot{\rho}_{R} =& -i[H_{rot},\rho_{R}]  +  \gamma_{h}(n_{h}+1)(e^{-i\omega_h\chi_h}b_{h}{\rho}_R b^{\dagger}_{h} - \frac{1}{2}\{b^{\dagger}_{h}     b_{h},{\rho}_R\})
    +  \gamma_{h}n_{h}( e^{i\omega_h\chi_h}b^{\dagger}_{h}{\rho}_R b_{h} 
    - \frac{1}{2}\{b_{h}  b^{\dagger}_{h},{\rho}_R\})\\
    &+\gamma_{c}(n_{c}+1)(e^{-i\omega_c\chi_c}b_{c}{\rho}_R b^{\dagger}_{c} - \frac{1}{2}\{b^{\dagger}_{c}  b_{c},{\rho}_R\})
    +  \gamma_{c}n_{c}( e^{i\omega_c\chi_c}b^{\dagger}_{c}{\rho}_R b_{c} - \frac{1}{2}\{b_{c}b^{\dagger}_{c},{\rho}_R\}). \nonumber
\end{align}
where $H_{rot}= -\delta \op{1} +\alpha(\ketbra{1}{0}+\ketbra{0}{1})$ and for the resonant driving case, we consider $\delta =0$ (see Appendix~\ref{ROT}).
Accordinlgly, the full Liouvillian super-operator $\mathcal{L}(\chi_{h},\chi_{c})$ with counting fields is
\begin{equation*}
 \left(
\begin{array}{ccccccccc}
 -\text{g}_{1}-\text{g}_{3} & 0 & 0 & 0 & \text{g}_{4} e^{i\chi_c\omega_c} & 0 & 0 & 0 & \text{g}_{2} e^{i \text{$\chi $}_h \text{$\omega $}_h} \\
 0 & -\frac{\text{g}_{1}}{2}-\frac{\text{g}_{3}}{2}-\frac{\text{g}_{4}}{2} & -i \alpha  & 0 & 0 & 0 & 0 & 0 & 0 \\
 0 & -i \alpha  & -\frac{\text{g}_1}{2}-\frac{\text{g}_2}{2}-\frac{\text{g}_3}{2} & 0 & 0 & 0 & 0 & 0 & 0 \\
 0 & 0 & 0 & -\frac{\text{g}_1}{2}-\frac{\text{g}_3}{2}-\frac{\text{g}_4}{2} & 0 & 0 & i \alpha  & 0 & 0 \\
 \text{g}_3 e^{-i\chi_c\omega_c} & 0 & 0 & 0 & -\text{g}_4 & -i \alpha  & 0 & i \alpha  & 0 \\
 0 & 0 & 0 & 0 & -i \alpha  & -\frac{\text{g}_2}{2}-\frac{\text{g}_4}{2} & 0 & 0 & i \alpha  \\
 0 & 0 & 0 & i \alpha  & 0 & 0 & -\frac{\text{g}_1}{2}-\frac{\text{g}_2}{2}-\frac{\text{g}_3}{2} & 0 & 0 \\
 0 & 0 & 0 & 0 & i \alpha  & 0 & 0 & -\frac{\text{g}_2}{2}-\frac{\text{g}_4}{2} & -i \alpha  \\
 \text{g}_1 e^{-i \text{$\chi $}_h \text{$\omega $}_h} & 0 & 0 & 0 & 0 & i \alpha  & 0 & -i \alpha  & -\text{g}_{2} \\ 
\end{array}
\right),
\end{equation*}
where $g_{1}=\gamma_{h}(n_{h}+1)$, $g_{2}=\gamma_{h}n_{h}$, $g_{3}=\gamma_{c}(n_{c}+1)$ and $g_{4}=\gamma_{c}n_{c}$. We set $\chi_{h}=\chi_{c}=\chi$ to calculate the power statistics. Following the previous discussion in this section, we find the polynomial factors with respective derivatives:
\begin{align*}
    a_1= &-\frac{1}{64} ({\gamma }_{c} {n}_{c}+{\gamma }_{h} {n}_{h}) \left(4 \alpha ^2+({\gamma}_{c}+{\gamma}_{h}+2 {\gamma}_{c} {n}_c +{\gamma}_{h} {n}_{h}) ({\gamma }_{c}+{\gamma_{h}}+{\gamma }_{c} n_{c}+2 {\gamma }_{h} n_{h})\right)^2 \left(4 \alpha ^2 ({\gamma }_{c} (3 n_{c}+2)+{\gamma }_{h} (3 n_{h}+2))
 \right. \\ & \left.+{\gamma }_{c} {\gamma }_{h} (3 n_{c}n_{h} +n_{c}+n_{h}) ({\gamma }_{c} n_{c}+{\gamma }_{h} n_{h})\right), \\
a_2=& -\frac{1}{64} (4 \alpha ^2+({\gamma }_{c}+{\gamma }_{h}+2 {\gamma }_{c}n_{c} +{\gamma }_{h}n_{h} ) ({\gamma }_{c}+{\gamma }_{h}+{\gamma }_{c}n_{c}+2 {\gamma }_{h}n_{h} )) ({\gamma }_{c}^5 n_{c}^2 (n_{c}+1) (2 n_{c}+1)^2+{\gamma }_{c}^2 {\gamma }_{h}^3 (5 n_{c}^2  
+(n_{c}\\
&(238 n_{c}+157)+25) n_{h}^3+(6 n_{c} (39 n_{c}+19)+11) n_{h}^2+n_{c} (67 n_{c}+18) n_{h})+{\gamma }_{c}^3 {\gamma }_{h}^2 (n_{c}^3 (n_{h} (238 n_{h} +157)+25)  
+n_{c}^2 \\
&(6 n_{h} (39 n_{h}+19)+11)+n_{c} n_{h}  (67 n_{h} +18)+5 n_{h}^2)+{\gamma }_{c} {\gamma }_{h}^4 n_{h} (n_{c} (n_{h} (n_{h} (82 n_{h}+113)+47)+6)+n_{h} (28 n_{h}^2+30 n_{h} \\&+7))+64 \alpha ^4 ({\gamma }_{c}+{\gamma }_{h}+2 {\gamma }_{c}n_{c} +2 n_{h}n_{h})+4 \alpha ^2 (({\gamma }_{c}^3 (3 n_{c}+2)^2 (6 n_{c}+1)+({\gamma }_{c}^2 ({\gamma }_{h} (n_{c} (2 (91 n_{c}+85) n_{h}+85 n_{c}+72) \\
&+36 n_{h}+12)+({\gamma }_{c}({\gamma }_{h}^2 (2 n_{c}  (n_{h} (91 n_{h}+85)+18)+n_{h} (85 n_{h}+72)+12)+({\gamma }_{h}^3 (3 n_{h}+2)^2 (6 n_{h}+1))+({\gamma }_{c}^4 ({\gamma }_{h} n_{c} \\
&(n_{c} (n_{c} (n_{c} (82 n_{h}+28)+113 n_{h}+30)+47 n_{h}+7)+6n_{h} )+({\gamma }_{h}^5 n_{h}^2 (n_{h}+1) (2 n_{h}+1)^2), \\  
a'_0=& \frac{1}{16} {\gamma }_{c} {\gamma }_{h} (n_{c}-n_{h}) (\omega_h-\omega_c) ({\gamma }_{c} n_{c}+{\gamma }_{h} n_{h}) \left(4 \alpha ^3+\alpha ({\gamma }_{c} +{\gamma }_{h}+2{\gamma }_{c}   n_{c}+{\gamma }_{h}  n_{h}) ({\gamma }_{c}+{\gamma }_{h}+{\gamma }_{c} n_{c}+2 {\gamma }_{h} n_{h} )\right)^2, \\
a''_0=& \frac{1}{16} \alpha ^2 {\gamma }_{c}{\gamma }_{h}  (2 n_{c} n_{h}+n_{c}+n_{h}) ({\omega}_c-{\omega}_h)^2 ({\gamma }_{c} n_{c}+{\gamma }_{h} n_{h}) \left(4 \alpha ^2+({\gamma }_{c}+{\gamma }_{h}+2 {\gamma }_{c}n_{c}+{\gamma }_{h} n_{h}) ({\gamma }_{c}+{\gamma }_{h}+{\gamma }_{c} n_{c} +2 {\gamma }_{h} n_{h})\right)^2, \\
\mbox{and} \ \ \ a'_1=&\frac{1}{8} \alpha ^2 {\gamma }_{c}{\gamma }_{h} (n_{c}-n_{h}) ({\omega }_{h}-{\omega }_{c}) \left(4 \alpha ^2+({\gamma }_{c})^2 (n_{c} (8 n_{c}+7)+1)+{\gamma }_{c}{\gamma }_{h}  (17 n_{c}n_{h} +7 n_{c}+7 n_{h}+2)+{\gamma }_{h}^2 (n_{h} (8 n_{h}+7)+1)\right) \\
& \ \ \ \ \ \ \ \left(4 \alpha ^2+({\gamma }_{c}+{\gamma }_{h}+2{\gamma }_{c} n_{c} +{\gamma }_{h} n_{h}) ({\gamma }_{c}+{\gamma }_{h}+{\gamma }_{c} n_{c}+2{\gamma }_{h}  n_{h})\right) 
\end{align*}

 Utilizing these expressions, the average power and the variance in power of IQHEs become
\begin{align}
\langle P_{I} \rangle &= -\frac{4 \alpha ^2 \gamma_{h}\gamma_{c}(n_{h}-n_{c}) }{4 \alpha ^2 (\gamma_{c} (3 n_{c}+2)+\gamma_{h} (3 n_{h}+2))+\gamma_{c}\gamma_{h}  (3  n_{c} n_{h} + n_{c}+n_{h})( \gamma_{c} n_{c} +\gamma_{h} n_{h})}(\omega_{h}-\omega_{c}), \\
\mbox{and} \ \ \ \Delta P_{I}&=(F_{p}|\langle P_{I}\rangle| - \frac{k}{\alpha^2(\omega_h-\omega_c)^2}|\langle P_{I}\rangle|^3  )(\omega_h-\omega_c),
\end{align}
where $F_{p}= \frac{2n_{h}n_{c}+n_{h}+n_{c}}{n_{h}-n_{c}}$ and $k =\frac{4\alpha^2}{\gamma_{0}^{2}(n_{h}-n_{c})}+\frac{n_{h}n_{c}+n_{c}^{2}+n_{h}^{2}}{n_{h}-n_{c}} +2F_{p}$. Now, the noise-to-signal ratio of the power of IQHEs is
\begin{equation}
   \mathcal{N}_{I}=  \frac{ \Delta P_{I}}{\langle P_{I} \rangle^2}=\left(\frac{F_{p}}{|\langle P_{I}\rangle| }- \frac{k}{\alpha^2(\omega_h-\omega_c)^2}|\langle P_{I}\rangle| \right)(\omega_h-\omega_c)=\frac{F_{p}}{\alpha \mathcal{C}(\sigma_{{I}})}\left(1- \frac{k}{F_{p}} {\mathcal{C}(\sigma_{{I}})^2}\right),
\end{equation}
where $\langle P_{I} \rangle =- \alpha (\omega_{h} - \omega_{c}) \mathcal{C}(\sigma_{I})$, and $\mathcal{C}(\sigma_{{I}})$ is $l$-1 norm of coherence of the steady state $\sigma_I$. It is important to note that the noise-to-signal ratio of currents, power, and photon number flux is the same for IQHEs.

\section{Classical thermodynamic uncertainty relation and power-efficiency-constancy trade-off relation}\label{CTUR}
Classical steady-state heat engines always exhibit trade-off relationships between relative fluctuation in output power, the thermodynamic cost (quantified by the rate of entropy production $\dot{S}$), and heat-to-work conversion efficiency. There are two trade-off relations
\begin{align}
\mathcal{Q} &=\dot{S} \frac{ \Delta P}{\langle P \rangle ^2} \geq 2\label{appeq:cTUR}, \\
\mbox{and} \
\mathcal{D}&  = ({\eta_{Cor}}-{\eta}) \frac{\rm \Delta P}{\langle P \rangle } \frac{\beta_{c} \omega_{h}} {(\omega_{h}-\omega_{c})}
 \geq 2 \label{appeq:PEC},
 \end{align}
where $\dot{S}$ the rate of entropy production, $\eta=1-\frac{\omega_{c}}{\omega_{h}}$ is the engine efficiency for both coherent and incoherent engines, and $\eta_{Cor}=1-\frac{\beta_{h}}{\beta_{c}}$ is the Carnot efficiency. Note, Eq.~\eqref{appeq:cTUR} is referred to as the classical thermodynamic uncertainty relation (cTUR)~\cite{Barato2015} and Eq.~\eqref{appeq:PEC} is referred to as the power-efficiency-constancy trade-off relation~\cite{Pietzonka2018}. 
The entropy production rate $\dot{S}$ for coherent and incoherent engines can be written as (for $X=C,I$)
\begin{equation}
  \dot{S}_{X}=- \beta_{h} \langle \dot{J}^{h}_{X} \rangle - \beta_{c}\langle \dot{J}_{X}^{c} \rangle= \ln{\left(\frac{n_{h}(n_{c}+1)}{n_{c}(n_{h}+1)}\right)}\langle \dot{N}_{X} \rangle > 0,
\end{equation}
where ${\langle\dot{N}_{X}\rangle}={|\langle{P}_{X}\rangle}|/(\omega_h-\omega_c)$ is the average photon number current, $\dot{J}^{X}_{h}$ and $\dot{J}^{X}_{c}$ are average heat currents corresponding hot and cold baths, respectively. Moreover, we can write 
\begin{equation}
     ({\eta_{Cor}}-{\eta}) \frac{\beta_{c} \omega_{h}} {(\omega_{h}-\omega_{c})}= \ln{\left(\frac{n_{h}(n_{c}+1)}{n_{c}(n_{h}+1)}\right)}.
\end{equation}
To obtain above expression we have used the relation $n_{x} = 1/(e^{\beta_{x}\omega_x}-1)$ for $x=h,c$. Using above relations, we can show that
\begin{equation}~\label{Unified}
    \mathcal{Q}_{X}= \mathcal{D}_{X} = \ln{\left(\frac{n_{h}(n_{c}+1)}{n_{c}(n_{h}+1)}\right)} F_{X}.
\end{equation}
Here $F_{X}=\frac{\Delta \dot{N}_X}{\langle\dot{N}_X\rangle}$ is known as the Fano factor of photon number current ($\dot{N}$), where ${\langle\dot{N}_{X}\rangle}={|\langle{P}_{X}\rangle}|/(\omega_h-\omega_c)$ and ${\Delta \dot{N}_{X}}=\Delta{P_{X}}/(\omega_{h}-\omega_{c})^2$ are variance and average of photon number current for the steady state dynamics. The Eq.~\eqref{Unified} indicates that in the context of CQHEs and IQHEs, both the cTUR and the power-efficiency-constancy trade-off relation coincide. By using the expression of $\langle P_X \rangle $ and $\Delta P_{X}$, the Fano factors for CQHEs and IQHEs can be respectively written in terms of population and energetic coherence as, 
\begin{align}
 F_{C}=F_{p}\left(1- \frac{3}{2}{(\mathcal{C}(\sigma_{{C}}))^2}\right),  \ \ \ 
 \mbox{and} \ \ \  F_{I}= F_{p}\left(1- \frac{k}{F_{p}} {(\mathcal{C}(\sigma_{{I}}))^2}\right).
 \end{align}

\section{Quantum Thermodynamic Uncertainty Relation}\label{QTURR}
A quantum formulation of the thermodynamic uncertainty relation was recently obtained for Markovian dynamics (described by the Lindblad master equation) using the quantum Cramér-Rao bound. The steady-state version of the QTUR reads~\cite{Hasegawa2020}:
\begin{equation}\label{qTURbound}
\mathcal{N} = \frac{\Delta P}{\langle P \rangle ^2} \geq f =\frac{1} {\Upsilon+ \Psi}.
\end{equation}
In the above bound~\eqref{qTURbound}, $\Upsilon$ denotes the quantum dynamical activity, which is the average rate of transitions in the steady-state and reads
\begin{equation}
 \Upsilon =   \sum_{k}\Tr(L_{k}^{\dagger}L_{k}\rho_{ss}),
\end{equation}
where $\rho_{ss}$ represent the steady state of the given system, $L_{k}$ and $L_{k}^{\dagger}$ represent the jump operators and its ad-joint operators, respectively. In the above bound~\eqref{qTURbound}, $\Psi$ denotes the coherent-dynamics contribution and reads
\begin{equation}
  \Psi = -4(\langle \langle \mathbb{I}|\mathcal{L}_{L}\mathcal{L}^{+}\mathcal{L}_{R}|\rho_{ss}\rangle\rangle+\langle \langle \mathbb{I}|\mathcal{L}_{R}\mathcal{L}^{+}\mathcal{L}_{L}|\rho_{ss}\rangle\rangle ),
\end{equation}
where $|\rho_{ss}\rangle\rangle$ denotes the vectorized steady-state density matrix $\rho_{ss}$, $|\mathbb{I}\rangle\rangle= \sum_{i}\ket{i}^{*} \otimes \ket{i}$ is the vectorized identity. $\mathcal{L}^{+}$ denotes the Drazin inverse of vectorized Liouvillian super operator ($\mathcal{L}=\mathcal{L}_{R}+\mathcal{L}_{L}$) and the expression of $\mathcal{L}_{R}$ and $\mathcal{L}_{L}$ reads as follows
\begin{align*}
    \mathcal{L}_{R}& = -i\mathcal{I} \otimes H+ \frac{1}{2}\sum_{k}(L_{k}^{*} \otimes L_{k}-\mathcal{I}\otimes L_{k}^{\dagger}L_{k} ), \\
 \mbox{and} \\
    \mathcal{L}_{L}& = iH^{T} \otimes \mathcal{I} +\frac{1}{2}\sum_{k}(L_{k}^{*} \otimes L_{k}- (L_{k}^{\dagger}L_{k})^{T} \otimes \mathcal{I}),
\end{align*}
where $H$ is the Hamiltonian of the system and $\mathcal{I}$ is the identity matrix. The vectorized Liouvillian super operator can be written as $\mathcal{L} = \sum_{j\neq 0}\lambda_{j} |x_{j}\rangle\rangle\langle \langle y_{j}|$, where $|x_{j}\rangle\rangle$ and  $|y_{j}\rangle\rangle$ are right and left eigenvectors of vectorized Liouvillian super operator, respectively and $\lambda_{j}$ is eigen value of vectorized Liouvillian super operator. The Drazin inverse of the Liouvillian super operator can be obtained by inverting the eigen values $\mathcal{L}^{+} = \sum_{j\neq 0}\frac{1}{\lambda_{j}} |x_{j}\rangle\rangle\langle \langle y_{j}|$~\cite{Landi2024}. The Drazin inverse also can be calculated using some alternative methods, for more details see Ref.~\cite{Landi2024}. Employing this definition, we derived the Drazin inverse of vectorized Liouvillian superoperators for CQHEs and IQHEs as
\begin{align*}
    \mathcal{L}^{+}_{C} = \left(
\begin{array}{cccc}
 \frac{4 \alpha ^2 (\gamma_1-3 \gamma_2)-\gamma_1 (\gamma_1+\gamma_2)^2}{\left(8 \alpha ^2+(\gamma_1+\gamma_2)^2\right)^2} & \frac{2 i \alpha }{8 \alpha ^2+(\gamma_1+\gamma_2)^2} & -\frac{2 i \alpha }{8 \alpha ^2+(\gamma_1+\gamma_2)^2} & \frac{4 \alpha ^2 (3 \gamma_1-\gamma_2)+\gamma_2 (\gamma_1+\gamma_2)^2}{\left(8 \alpha ^2+(\gamma_1+\gamma_2)^2\right)^2} \\
 \frac{4 i \alpha  \left(4 \alpha ^2+(2 \gamma_1-\gamma_2) (\gamma_1+\gamma_2)\right)}{\left(8 \alpha ^2+(\gamma_1+\gamma_2)^2\right)^2} & -\frac{2 \left(4 \alpha ^2+(\gamma_1+\gamma_2)^2\right)}{(\gamma_1+\gamma_2) \left(8 \alpha ^2+(\gamma_1+\gamma_2)^2\right)} & -\frac{8 \alpha ^2}{(\gamma_1+\gamma_2) \left(8 \alpha ^2+(\gamma_1+\gamma_2)^2\right)} & -\frac{4 i \alpha  \left(4 \alpha ^2-(\gamma_1-2 \gamma_2) (\gamma_1+\gamma_2)\right)}{\left(8 \alpha ^2+(\gamma_1+\gamma_2)^2\right)^2} \\
 -\frac{4 i \alpha  \left(4 \alpha ^2+(2 \gamma_1-\gamma_2) (\gamma_1+\gamma_2)\right)}{\left(8 \alpha ^2+(\gamma_1+\gamma_2)^2\right)^2} & -\frac{8 \alpha ^2}{(\gamma_1+\gamma_2) \left(8 \alpha ^2+(\gamma_1+\gamma_2)^2\right)} & -\frac{2 \left(4 \alpha ^2+(\gamma_1+\gamma_2)^2\right)}{(\gamma_1+\gamma_2) \left(8 \alpha ^2+(\gamma_1+\gamma_2)^2\right)} & \frac{4 i \alpha  \left(4 \alpha ^2-(\gamma_1-2 \gamma_2) (\gamma_1+\gamma_2)\right)}{\left(8 \alpha ^2+(\gamma_1+\gamma_2)^2\right)^2} \\
 \frac{\gamma_1 (\gamma_1+\gamma_2)^2-4 \alpha ^2 (\gamma_1-3 \gamma_2)}{\left(8 \alpha ^2+(\gamma_1+\gamma_2)^2\right)^2} & -\frac{2 i \alpha }{8 \alpha ^2+(\gamma_1+\gamma_2)^2} & \frac{2 i \alpha }{8 \alpha ^2+(\gamma_1+\gamma_2)^2} & \frac{4 \alpha ^2 (\gamma_2-3 \gamma_1)-\gamma_2 (\gamma_1+\gamma_2)^2}{\left(8 \alpha ^2+(\gamma_1+\gamma_2)^2\right)^2} \\
\end{array}
\right)
\end{align*}
and 
\begin{align*}
    \mathcal{L}^{+}_{I} = \left(
    \begin{array}{ccccccccc}
    a_{11} & 0 & 0 & 0 & a_{15} & a_{16} & 0 & a_{18} & a_{19} \\
    0 & a_{22} & a_{23} & 0 & 0 & 0 & 0 & 0 & 0 \\
    0 & a_{32} & a_{33} & 0 & 0 & 0 & 0 & 0 & 0 \\
  0 & 0 & 0 & a_{44} & 0 & 0 & a_{47} & 0 & 0 \\
    a_{51} & 0 & 0 & 0 & a_{55} & a_{56} & 0 & a_{58} & a_{59} \\
    a_{61} & 0 & 0 & 0 & a_{65} & a_{66} & 0 & a_{68} & a_{69} \\
    0 & 0 & 0 & a_{74} & 0 & 0 & a_{77} & 0 & 0 \\
    a_{81} & 0 & 0 & 0 & a_{85} & a_{86} & 0 & a_{88} & a_{89} \\
    a_{91} & 0 & 0 & 0 & a_{95} & a_{96} & 0 & a_{98} & a_{99} \\
    \end{array}
    \right),
\end{align*}
respectively, where
\begin{align*}
    &\gamma_1=\gamma_0n_c(n_h+1),\\
&\gamma_2=\gamma_0 n_h(n_c+1),
\end{align*}
\begin{align*}
&a_{11}=-\frac{4 \alpha ^2 \gamma_0 ^2 (n_{c}+n_{h}+2) \left(n_{c}^2+6 n_{c} n_{h}+n_{h}^2\right)+\gamma_0 ^4 (n_{c}+n_{h})^2 \left(n_{c}^2 (n_{h}+1)+n_{c} n_{h}^2+n_{h}^2\right)+64 \alpha ^4 (n_{c}+n_{h}+2)}{\gamma_0  \left(4 \alpha ^2 (3 n_{c}+3 n_{h}+4)+\gamma_0 ^2 (n_{c}+n_{h}) (3 n_{c} n_{h}+n_{c}+n_{h})\right)^2},\\
&a_{61}=-a_{81}=\frac{2 i \alpha  (n_{c}-n_{h}) \left(\gamma_0 ^2 \left(3 n_{c}^2 (n_{h}+1)+n_{c} (3 n_{h} (n_{h}+4)+4)+n_{h} (3 n_{h}+4)\right)+12 \alpha ^2 (n_{c}+n_{h}+2)\right)}{\left(4 \alpha ^2 (3 n_{c}+3 n_{h}+4)+\gamma_0 ^2 (n_{c}+n_{h}) (3 n_{c} n_{h}+n_{c}+n_{h})\right)^2},\\
    & a_{23}=a_{32}=-a_{74}=-a_{47}=\frac{4 i \alpha }{4 \alpha ^2+\gamma_0 ^2 (2 n_{c}+n_{h}+2) (n_{c}+2 n_{h}+2)},\\
     & a_{22}=a_{44}=-\frac{2 \gamma_0  (n_{c}+2 n_{h}+2)}{4 \alpha ^2+\gamma_0 ^2 (2 n_{c}+n_{h}+2) (n_{c}+2 n_{h}+2)},\\
    &  a_{33}=a_{77}   = -\frac{2 \gamma_0  (2 n_{c}+n_{h}+2)}{4 \alpha ^2+\gamma_0 ^2 (2 n_{c}+n_{h}+2) (n_{c}+2 n_{h}+2)},\\
     &a_{16}=-a_{18}=\frac{2 i \alpha  (n_{c}-n_{h})}{4 \alpha ^2 (3 n_{c}+3 n_{h}+4)+\gamma_0 ^2 (n_{c}+n_{h}) (3 n_{c} n_{h}+n_{c}+n_{h})},\\
 &a_{56}=-a_{58}=\frac{2 i \alpha  (n_{c}+2 n_{h}+2)}{4 \alpha ^2 (3 n_{c}+3 n_{h}+4)+\gamma_0 ^2 (n_{c}+n_{h}) (3 n_{c} n_{h}+n_{c}+n_{h})},\\
     &a_{96}=a_{98}=-\frac{2 i \alpha  (2 n_{c}+n_{h}+2)}{4 \alpha ^2 (3 n_{c}+3 n_{h}+4)+\gamma_0 ^2 (n_{c}+n_{h}) (3 n_{c} n_{h}+n_{c}+n_{h})},\\
&a_{86}=a_{68}=-\frac{4 \alpha ^2 (3 n_{c}+3 n_{h}+4)}{\gamma_0  (n_{c}+n_{h}) \left(4 \alpha ^2 (3 n_{c}+3 n_{h}+4)+\gamma_0 ^2 (n_{c}+n_{h}) (3 n_{c} n_{h}+n_{c}+n_{h})\right)},\\
    &a_{66}=a_{88}=\frac{-4 \alpha ^2 (3 n_{c}+3 n_{h}+4)-2 \gamma_0 ^2 (n_{c}+n_{h}) (3 n_{c} n_{h}+n_{c}+n_{h})}{\gamma_0  (n_{c}+n_{h}) \left(4 \alpha ^2 (3 n_{c}+3 n_{h}+4)+\gamma_0 ^2 (n_{c}+n_{h}) (3 n_{c} n_{h}+n_{c}+n_{h})\right)},\\
    &a_{15}=\frac{4 \alpha ^2 \gamma_0 ^2 \left(-n_{c}^3+(5 n_{c}+4) n_{h}^2+2 (n_{c}-2) n_{c} n_{h}+2 n_{h}^3\right)-\gamma_0 ^4 n_{c} (n_{c}+n_{h})^2 \left((n_{c}-1) n_{h}+n_{c}-2 n_{h}^2\right)+32 \alpha ^4 (n_{c}+n_{h})}{\gamma_0  \left(4 \alpha ^2 (3 n_{c}+3 n_{h}+4)+\gamma_0 ^2 (n_{c}+n_{h}) (3 n_{c} n_{h}+n_{c}+n_{h})\right)^2},\\
      &a_{19}=\frac{4 \alpha ^2 \gamma_0 ^2 \left(2 n_{c}^3+n_{c}^2 (5 n_{h}+4)+2 n_{c} (n_{h}-2) n_{h}-n_{h}^3\right)+32 \alpha ^4 (n_{c}+n_{h})+\gamma_0 ^4 n_{h} (n_{c}+n_{h})^2 (n_{c} (2 n_{c}+1)-(n_{c}+1) n_{h})}{\gamma_0  \left(4 \alpha ^2 (3 n_{c}+3 n_{h}+4)+\gamma_0 ^2 (n_{c}+n_{h}) (3 n_{c} n_{h}+n_{c}+n_{h})\right)^2},\\
    &a_{65}=-a_{85}=\frac{2 i \alpha  \left(4 \alpha ^2 \left(3 n_{c}^2+3 n_{c} (3 n_{h}+4)+6 n_{h} (n_{h}+2)+8\right)+\gamma_0 ^2 \left(3 n_{c}^3 (n_{h}+1)+3 n_{c}^2 (3 n_{h} (n_{h}+2)+2)+n_{c} n_{h} \left(6 n_{h}^2+3 n_{h}+4\right)-2 n_{h}^2\right)\right)}{\left(4 \alpha ^2 (3 n_{c}+3 n_{h}+4)+\gamma_0 ^2 (n_{c}+n_{h}) (3 n_{c} n_{h}+n_{c}+n_{h})\right)^2},\\
& a_{95}=\frac{4 \alpha ^2 \gamma_0 ^2 \left(n_{c}^2 (2 n_{c}+3)+(5 n_{c}+7) n_{h}^2+(n_{c}+2) (5 n_{c}+4) n_{h}+2 n_{h}^3\right)-16 \alpha ^4 (n_{c}+n_{h})+2 \gamma_0 ^4 n_{c} (n_{h}+1) (n_{c}+n_{h})^2 (n_{c}+n_{h}+1)}{\gamma_0  \left(4 \alpha ^2 (3 n_{c}+3 n_{h}+4)+\gamma_0 ^2 (n_{c}+n_{h}) (3 n_{c} n_{h}+n_{c}+n_{h})\right)^2},\\
& a_{59}=\frac{4 \alpha ^2 \gamma_0 ^2 \left(2 n_{c}^3+n_{c}^2 (5 n_{h}+7)+n_{c} (n_{h}+2) (5 n_{h}+4)+n_{h}^2 (2 n_{h}+3)\right)-16 \alpha ^4 (n_{c}+n_{h})+2 \gamma_0 ^4 (n_{c}+1) n_{h} (n_{c}+n_{h})^2 (n_{c}+n_{h}+1)}{\gamma_0  \left(4 \alpha ^2 (3 n_{c}+3 n_{h}+4)+\gamma_0 ^2 (n_{c}+n_{h}) (3 n_{c} n_{h}+n_{c}+n_{h})\right)^2},\\
& a_{89}=-a_{69}=\frac{2 i \alpha  \left(4 \alpha ^2 \left(6 n_{c}^2+3 n_{c} (3 n_{h}+4)+3 n_{h} (n_{h}+4)+8\right)+\gamma_0 ^2 \left(6 n_{c}^3 n_{h}+n_{c}^2 \left(9 n_{h}^2+3 n_{h}-2\right)+n_{c} n_{h} (3 n_{h} (n_{h}+6)+4)+3 n_{h}^2 (n_{h}+2)\right)\right)}{\left(4 \alpha ^2 (3 n_{c}+3 n_{h}+4)+\gamma_0 ^2 (n_{c}+n_{h}) (3 n_{c} n_{h}+n_{c}+n_{h})\right)^2},\\
& a_{99}= -\frac{4 \alpha ^2 \gamma_0 ^2 \left((7 n_{c}+3) n_{h}^2+10 n_{c} (n_{c}+1) n_{h}+n_{c} (n_{c} (4 n_{c}+11)+8)+n_{h}^3\right)+16 \alpha ^4 (n_{c}+n_{h})+\gamma_0 ^4 n_{h} (n_{c}+n_{h})^2 (n_{c} (4 n_{c}+n_{h}+5)+n_{h}+2)}{\gamma_0  \left(4 \alpha ^2 (3 n_{c}+3 n_{h}+4)+\gamma_0 ^2 (n_{c}+n_{h}) (3 n_{c} n_{h}+n_{c}+n_{h})\right)^2},\\
& a_{55}=-\frac{4 \alpha ^2 \gamma_0 ^2 \left(n_{c}^3+n_{c}^2 (7 n_{h}+3)+10 n_{c} n_{h} (n_{h}+1)+n_{h} (n_{h} (4 n_{h}+11)+8)\right)+\gamma_0 ^4 n_{c} (n_{c}+n_{h})^2 \left((n_{c}+5) n_{h}+n_{c}+4 n_{h}^2+2\right)+16 \alpha ^4 (n_{c}+n_{h})}{\gamma_0  \left(4 \alpha ^2 (3 n_{c}+3 n_{h}+4)+\gamma_0 ^2 (n_{c}+n_{h}) (3 n_{c} n_{h}+n_{c}+n_{h})\right)^2},
 \end{align*}

\begin{align*}
&a_{51}=\\
 &\frac{4 \alpha ^2 \gamma_0 ^2 \left(n_{c} \left(n_{c}-n_{c}^2+4\right)+(5 n_{c}+1) n_{h}^2+2 n_{c} (n_{c}+3) n_{h}+2 n_{h}^3-4 n_{h}\right)-\gamma_0 ^4 (n_{c}+1) (n_{c}+n_{h})^2 \left((n_{c}-1) n_{h}+n_{c}-2 n_{h}^2\right)+32 \alpha ^4 (n_{c}+n_{h}+2)}{\gamma_0  \left(4 \alpha ^2 (3 n_{c}+3 n_{h}+4)+\gamma_0 ^2 (n_{c}+n_{h}) (3 n_{c} n_{h}+n_{c}+n_{h})\right)^2},
 \end{align*}
 and 
 \begin{align*}
&a_{91}= \\
    &\frac{4 \alpha ^2 \gamma_0 ^2 \left(2 n_{c}^3+n_{c}^2 (5 n_{h}+1)+2 n_{c} (n_{h} (n_{h}+3)-2)+n_{h} \left(n_{h}-n_{h}^2+4\right)\right)+32 \alpha ^4 (n_{c}+n_{h}+2)+\gamma_0 ^4 (n_{h}+1) (n_{c}+n_{h})^2 (n_{c} (2 n_{c}+1)-(n_{c}+1) n_{h})}{\gamma_0  \left(4 \alpha ^2 (3 n_{c}+3 n_{h}+4)+\gamma_0 ^2 (n_{c}+n_{h}) (3 n_{c} n_{h}+n_{c}+n_{h})\right)^2}.
\end{align*}
The superoperators $\mathcal{L}_{R}$ and $\mathcal{L}_{L}$ for CQHEs and IQHEs can be computed using the corresponding jump operators {$\sqrt{\gamma_0 n_{c}(n_{h}+1)} b_{h}$, $\sqrt{\gamma_0 n_{h}(n_{c}+1)} b_{hc}^{\dagger}$} and {$\sqrt{\gamma_0(n_{h}+1)} b_{h}$, $\sqrt{\gamma_0 n_{h}} b_{h}^{\dagger}$, $\sqrt{\gamma_0(n_{c}+1)} b_{c}$, $\sqrt{\gamma_0 n_{c}} b_{c}^{\dagger}$} through a simple exercise. The expressions of the lower bounds ($f_X$) on the noise-to-signal ratio of power for CQHEs and IQHEs in terms of driving and bath parameters are as follows
 \begin{align*}
  \frac{1}{f_C} &=  \frac{ 2\left(2 \alpha ^2+\gamma_0 ^2{n_h} {n_c} ({n_c}+1)  ({n_h}+1)\right) \left(32 \alpha ^2+\gamma_{0} ^2 (2 {n_c} {n_h}+{n_c}+{n_h})^2\right)}{ \gamma_{0}(n_h + n_c + 2 n_h n_c) (8 \alpha^{2}  + \gamma_0^{2}(n_h + n_c + 2 n_h n_c)^2) },\\
 \ \ \ 
 \mbox{and} \ \ \ 
  \frac{1}{f_I} & = \frac{2 (n_h + n_c+2) (  4 \alpha^2+\gamma_0^2n_h n_c ) ( 16 \alpha^2+ \gamma_0^2 (n_h + n_c)^2)}{\gamma_0(n_h + n_c)  (4 \alpha^2(4 + 3 (n_h +  n_c))+\gamma_0^{2}(n_h + n_c) (n_h + n_c + 3 n_h n_c)   )}.
\end{align*}
It is important to note that the noise-to-signal ratio of currents, power, and photon number flux is the same for CQHEs as well as for IQHEs.

\section{Comparison between coherent and incoherent quantum heat engines for non-resonant driving}\label{NonReso}
In the main text and previous sections, we compared the coherent and incoherent heat engines for the resonant driving case $\delta=0$. In this section, we compare the coherent and incoherent heat engines for the non-resonant driving case, i.e., $\delta=\omega_{d}-(\omega_h-\omega_c)\neq0$ (as considered in  Ref.~\cite{Kalaee2021}). In this case, the total Hamiltonian of both engines in the rotating frame can be written as (see Appendix~\ref{ROT})
In the main text and previous sections, we compared the coherent and incoherent heat engines for the resonant driving case $\delta=0$. In this section, we compare the coherent and incoherent heat engines for the non-resonant driving case, i.e., $\delta=\omega_{d}-(\omega_h-\omega_c)\neq0$ (as considered in  Ref.~\cite{Kalaee2021}). In this case, the total Hamiltonian of both engines in the rotating frame can be written as (see Appendix~\ref{ROT})
\begin{equation}
    H_{rot} = -\delta \op{1} + \alpha( |1 \rangle \langle 0| + |0 \rangle \langle 1|)
\end{equation}
where detuning parameter $\delta=\omega_{d}-(\omega_{h}-\omega_{c})$. In this case, the steady state of the incoherent and coherent heat engine is given as (obtained by solving Eq.~\eqref{IQHER} and Eq.~\eqref{CQHERot} with $\delta\neq0$)
\begin{align}
   & \sigma_{I} = \frac{1}{\Theta_1}\{((g_{2}+g_{4}) \left(g_{1} g_{4} (g_{2}+g_{4})+4 \alpha ^2 (g_{1}+g_{3})\right)+4 \delta ^2 g_{1} g_{4} ) \op{0}  - 2 i \alpha  (2 i \delta +g_{2}+g_{4}) (g_{2} g_{3}-g_{1} g_{4})  \ketbra{0}{1} \nonumber\\ 
 &+ 2 i \alpha  (-2 i \delta +g_{2}+g_{4}) (g_{2} g_{3}-g_{1} g_{4}) \ketbra{1}{0} + ( (g_{2}+g_{4}) \left(4 \alpha ^2 (g_{1}+g_{3})+g_{2} g_{3} (g_{2}+g_{4})\right)+4 \delta ^2 g_{2} g_{3})  \op{1} \nonumber  \\
&+ ((g_{2}+g_{4})^2 \left(4 \alpha ^2+g_{2} g_{4}\right)+4 \delta ^2 g_{2} g_{4})\op{2}\} \\ 
&\ \mbox{and} \ \nonumber\\
&\sigma_{C} = \frac{1}{\Theta_2}\{ {4 \alpha ^2 (\gamma_1+\gamma_2)+\gamma_1 \left((\gamma_1+\gamma_2)^2+4 \delta ^2\right)}\op{0}  + {2 i \alpha  (\gamma_1-\gamma_2) (\gamma_1+\gamma_2-2 i \delta )} \ketbra{0}{1} \nonumber \\
  &   - {2 i \alpha  (\gamma_1-\gamma_2) (\gamma_1+\gamma_2+2 i \delta )} \ketbra{1}{0} + {4 \alpha ^2 (\gamma_1+\gamma_2)+\gamma_2 \left((\gamma_1+\gamma_2)^2+4 \delta ^2\right)}\op{1}\}.
 \end{align}

where 
where $\Theta_1=(g_{2}+g_{4}) \left(4 \alpha ^2 (2 g_{1}+g_{2}+2 g_{3}+g_{4})+(g_{2}+g_{4}) (g_{1} g_{4}+g_{2} (g_{3}+g_{4}))\right)+4 \delta ^2 (g_{1} g_{4}+g_{2} (g_{3}+g_{4}))$, $\Theta_{2}={(\gamma_1+\gamma_2) \left(8 \alpha ^2+(\gamma_1+\gamma_2)^2+4 \delta ^2\right)}$, $g_{1}=\gamma_{0}(n_{h}+1)$, $g_{2}=\gamma_{0}n_{h}$, $g_{3}=\gamma_{0}(n_{c}+1)$, $g_{4}=\gamma_{0}n_{c}$, $\gamma_1=\gamma_0 n_c(n_h+1)$ and $\gamma_2=\gamma_0n_h(n_c+1)$.

In general, for non-resonant driving, the average power and fluctuation in power of coherent and incoherent heat engines can be written as (for $X=I, C$)
\begin{equation}
   \langle P_{X} \rangle = -  \langle \dot{N}_{X} \rangle (\omega_{h}-\omega_{c})\ \mbox{and} \    \Delta P_{X} = \Delta \dot{N}_{X}  (\omega_{h}-\omega_{c})^2,
\end{equation}
 where $\langle \dot{N}_{X} \rangle=2\alpha \Im{\rho_{ij}^{X}}$. The expressions of average and variance of photon flux for both engines can be written as
\begin{equation}
   \langle \dot{N}_{C} \rangle= \frac{4 \alpha^2    (\gamma_2-\gamma_1)}{8 \alpha ^2+(\gamma_1+\gamma_2)^2+4 \delta ^2} \ \mbox{and} \ \Delta \dot{N}_{C} =
\langle \dot{N}_{C} \rangle \left[F_P - \frac{1}{2 \alpha^2 (\gamma_2^2 - \gamma_1^2)} \left( 3 (\gamma_1 + \gamma_2)^2 - 4\delta^2  \right) \langle \dot{N}_{C} \rangle^2\right]
\end{equation}

\begin{align}
&\langle \dot{N}_{I} \rangle=  \frac{4 \alpha^2  (g_{2}+g_{4}) (g_{2} g_{3}-g_{1} g_{4})}{(g_{2}+g_{4}) \left(4 \alpha ^2 (2 g_{1}+g_{2}+2 g_{3}+g_{4})+(g_{2}+g_{4}) (g_{1} g_{4}+g_{2} (g_{3}+g_{4}))\right)+4 \delta ^2 (g_{1} g_{4}+g_{2} (g_{3}+g_{4}))} \nonumber\\
&\ \mbox{and} \ \Delta \dot{N}_{I} =
\langle \dot{N}_{I} \rangle \left(F_p - \frac{k^{'}}{\alpha^2} \langle \dot{N}_{I} \rangle^2 \right).
\end{align}
 where $F_{p}= \frac{2n_{h}n_{c}+n_{h}+n_{c}}{n_{h}-n_{c}}$ and $k^{'} =\frac{4\alpha^2}{\gamma_{0}^{2}(n_{h}-n_{c})}+\frac{n_{h}n_{c}+n_{c}^{2}+n_{h}^{2}}{n_{h}-n_{c}} +2F_{p}+\frac{4 (n_c^2 - n_c n_h + n_h^2) \delta^2}{ \gamma_0^2(n_h - n_c) (n_c + n_h)^2 }$.

It is important to note that the above expression of average and variance of photon flux are determined using full-counting statistics like the resonant case (see section~\ref{appsec:FCS}). Using the above expressions, the noise-to-signal ratio of the power of IQHEs and CQHEs for non-resonant case can be written, respectively, as

\begin{equation}
\mathcal{N}_{C} =
\frac{F_P}{\langle \dot{N}_{C} \rangle} \left(1 - \frac{3 (\gamma_1 + \gamma_2)^2 - 4\delta^2}{2 \alpha^2 (\gamma_2^2 - \gamma_1^2)F_P} \langle \dot{N}_{C} \rangle^2\right)\ \mbox{and} \ \mathcal{N}_{I} =\frac{F_p}{\langle \dot{N}_{I} \rangle} \left(1 - \frac{k^{'}}{\alpha^2F_{p}} \langle \dot{N}_{I} \rangle^2 \right).
\end{equation}

The expression of $k'$ can be rewritten as
\begin{equation}
k'= \frac{2\alpha^2}{\langle \dot{N}_{I} \rangle} \left[\left( \frac{4 \alpha^2 \Gamma}{\delta^2 + \Gamma^2} + 4 \Gamma + 2 \gamma_0 \right) \frac{1}{K} + \left( \frac{\Gamma^2 - \delta^2}{\delta^2 + \Gamma^2} \right) \left( \frac{\gamma_0^2}{\Gamma} \right) \left( \frac{3 n_h n_c + n_h + n_c}{K} \right) \right]
\end{equation}
where
$K = \gamma_0^2 (3 n_h n_c + n_c + n_h) +  \frac{4 \alpha^2 \Gamma}{\delta^2 + \Gamma^2} (3 \Gamma + 2 \gamma_0)$ and $\Gamma =\frac{\gamma_0}{2} (n_c + n_h)$.

\begin{figure*}
    \centering
\includegraphics[width=8.8cm]{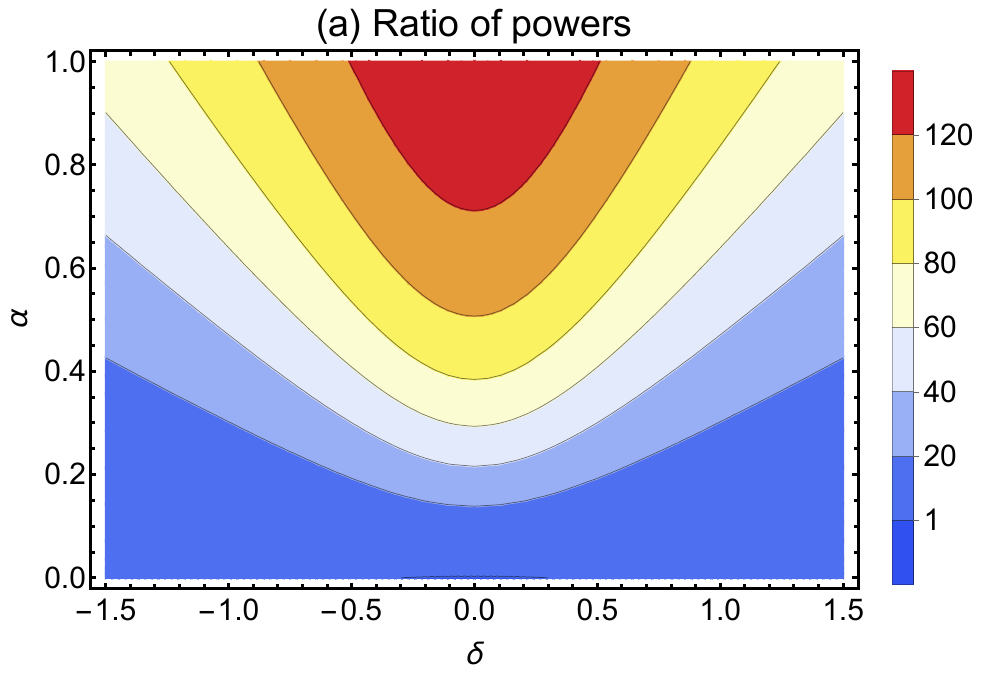}
    \space
\space
\space
\includegraphics[width=8.8cm]{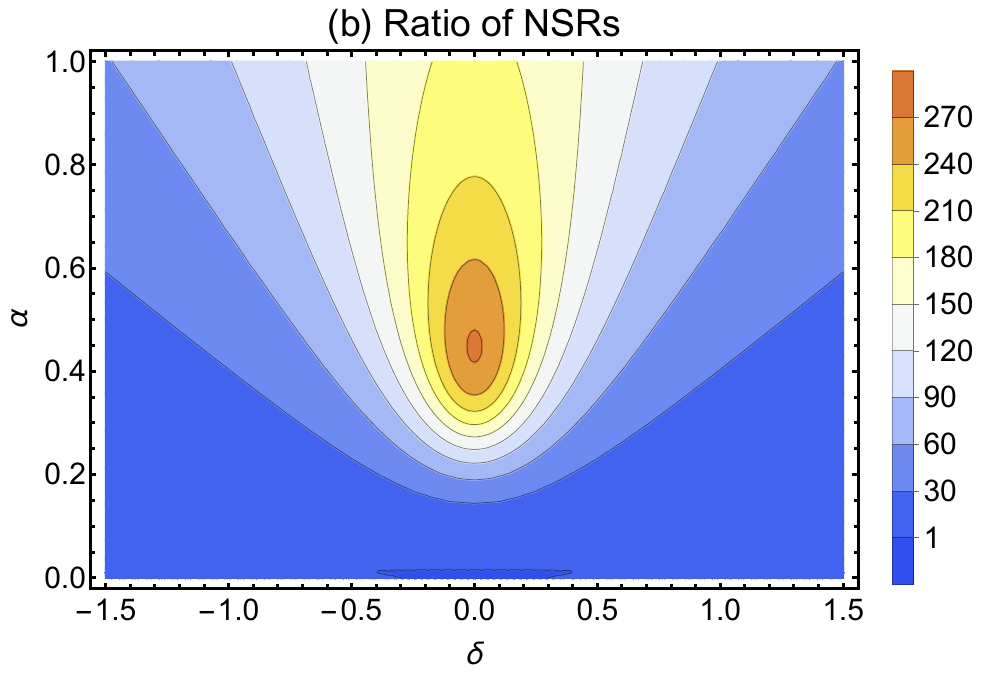}
\caption{{\bf Comparisons of ratio powers and noise-to-signal ratios (NSRs) in coherent and incoherent engines.} The computations are carried out with the parameters $\gamma_{0} = 0.01$, $\omega_{h} = 10$, $\omega_{c} = 5$, $\beta_h=0.8$ and $\beta_h=0.001$. (a) The figure of the right displays the ratio of powers $\mathcal{P}_{C}/\mathcal{P}_{I} = \langle \dot{N}_{C} \rangle/ \langle \dot{N}_{I} \rangle$ of the coherent and incoherent heat engine, against $\alpha$ and $\delta$. In fact, for these parameters, the ratio can be always $\mathcal{P}_{C} /\mathcal{P}_{I} \geq 1$ except for very small $\alpha$. (b) The figure of the right displays the ratio of noise-to-signal ratios (NSRs) in powers  $\mathcal{N}_{I}/\mathcal{N}_{C}$ of the coherent and incoherent heat engine, against $\alpha$ and $\delta$. In fact, for these parameters, the ratio can be always $\mathcal{N}_{I}/\mathcal{N}_{C} \geq 1$ except for very small $\alpha$. }
    \label{fig:CoherenceandPower1}
\end{figure*}

In Fig.~\ref{fig:CoherenceandPower1}(a), we plot the ratio of the average power for both coherent and incoherent engines. We find that with arbitrary detuning parameter \(|\delta| \geq 0\), the coherent engine outperforms the incoherent engine, except for very small values of the driving parameter \(\alpha\).

The noise-to-signal ratios (NSRs) may increase or decrease with an increase in the average photon flux, depending on the threshold value of the detuning parameter $\delta$ for both engines. This can be seen from the expression of $\mathcal{N}_C$ (and $\mathcal{N}_I$), particularly the second term in the parenthesis. For the detuning parameter \(|\delta| \leq \sqrt{3} (\Gamma + \gamma_0 n_h n_c)\), the noise-to-signal ratio of power in coherent heat engines is suppressed as power (photon flux) increases. A similar behavior is observed for the value \(|\delta| \leq \Gamma\), in incoherent engines. However, for large detuning parameters, specifically with \(|\delta| > \sqrt{3} (\Gamma + \gamma_0 n_h n_c)\) for coherent heat engines and with \(|\delta| > \Gamma\) for incoherent heat engines, the sign of the second term in the parenthesis flips (changes from negative to positive), leading to an increase in the noise-to-signal ratio as power (photon flux) increases. It is important to note that the detuning threshold for coherent heat engines is higher than that for incoherent heat engines (i.e., $\sqrt{3} (\Gamma + \gamma_0 n_h n_c) > \Gamma $), making the coherent engine advantageous in comparison.  In Fig.~\ref{fig:CoherenceandPower1}(b), we plot the noise-to-signal ratio of power with respect to the detuning parameter and driving parameter \(\alpha\). Our results show that the coherent engine consistently outperforms the incoherent engine, except for very small values of the driving parameter \(\alpha\). 

Thus, we conclude that the coherent heat engine remains advantageous over the incoherent heat engine in both resonant and non-resonant driving cases due to coherent heat transfer. Note that the analysis of the noise-to-signal ratio for incoherent heat engines (which are the standard SSD heat engines) has been rigorously studied in Ref.~\cite{Kalaee2021} for the non-resonant case to demonstrate their performance over their classical counterpart.

\twocolumngrid

\bibliography{che}

\end{document}